# DYNAMIC ESTIMATES OF DISPLACEMENT IN DISASTER REGIONS

A policy-driven framework triangulating data

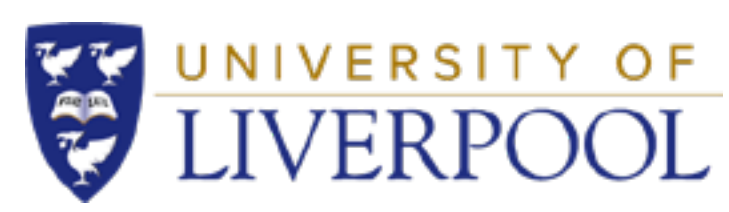

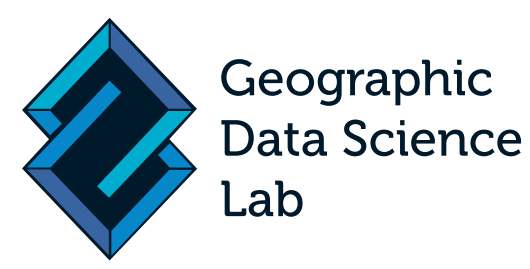

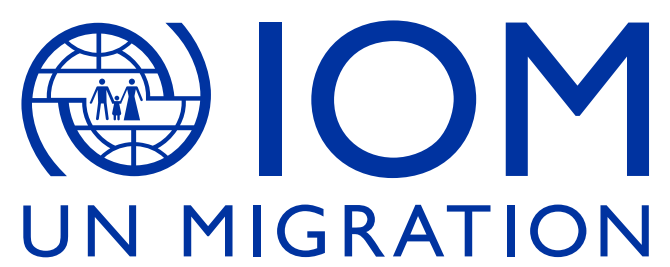

IOM is committed to the principle that humane and orderly migration benefits migrants and society. As an intergovernmental organization, IOM acts with its partners in the international community to assist in meeting the operational challenges of migration; advance understanding of migration issues; encourage social and economic development through migration; and uphold the human dignity and well-being of migrants.

Publisher: International Organization for Migration
17 route des Morillons
P.O. Box 17
1211 Geneva 19
Switzerland
Tel.: +41 22 717 9111
Fax: +41 22 798 6150
Email: hq@iom.int
Website: www.iom.int

This publication was issued without formal editing by IOM.

Cover image: Lea Riggi (2025)
Visual credits: Brian McDonald (2022)

Required citation: Pietrostefani, E., M. Mason, R. Iradukunda, H. Tran-Jones, I. Loktieva and F. Rowe (2025). *Dynamic Estimates of Displacement in Disaster Regions: A Policy-Driven Framework Triangulating Data*. IOM, Geneva.

ISBN 978-92-9278-079-1 (PDF)
DOI: https://doi.org/10.48550/arXiv.2511.01955

# DYNAMIC ESTIMATES OF DISPLACEMENT IN DISASTER REGIONS

A policy-driven framework triangulating data

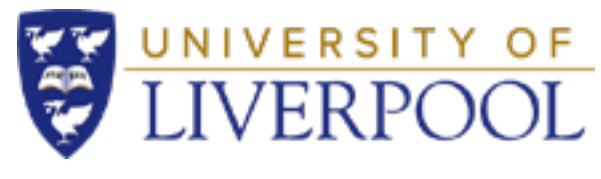

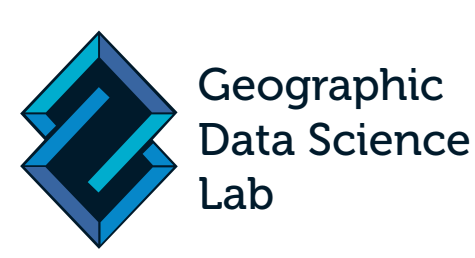

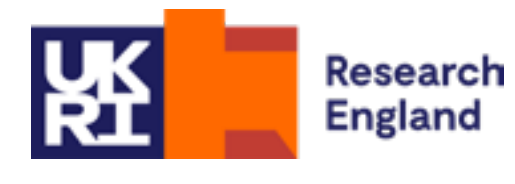

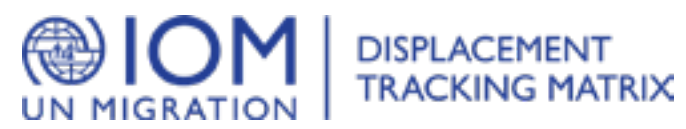

# FOREWORD

The global challenge of internal displacement, exacerbated by conflict, climate-induced natural hazards and disasters, requires innovative and collaborative approaches to ensure effective responses. At the end of 2024, an estimated 123.2 million people were forcibly displaced (UNHCR, 2025a). This staggering figure includes 3.8 million people uprooted within Ukraine due to war, 20.8 million internally displaced across the Horn of Africa by a combination of drought and violence and millions more affected annually in South and South-East Asia, where disasters such as cyclones triggered at least 1.8 million displacements in 2024 alone (ibid.). These estimates are a stark reminder of the need for timely, accurate and spatially detailed data and accessible ways to process it to inform humanitarian action and policy.

This publication represents a step forward in addressing the need for more integrated and adaptive displacement data systems. By triangulating traditional data sources with cutting-edge digital trace data – such as mobile phone GPS and social media data – it highlights key considerations for effectively combining these approaches in humanitarian contexts. The report explores these insights, through focused case studies, offer practical guidance for integrating diverse data streams to support more timely and informed interventions. The push for data innovation is particularly timely as funding cuts have reduced the humanitarian sector traditional data capabilities.

We demonstrate the power of triangulating data using two study cases. First, we use the first year of the war in Ukraine, which escalated in February 2022, and show the potential of data triangulation to generate actionable insights in complex and rapidly evolving settings that demand timely data, policy decisions and humanitarian responses. Second, we focus on the Pakistan floods of August 2022 to test the broader applicability of our approach – shifting the context from conflict to a climate-induced disaster setting. Together, the Ukraine and Pakistan case studies illustrate the versatility and effectiveness of data triangulation in supporting timely, data-driven responses across different types of crises.

This project has been made possible through the collaborative efforts of the Geographic Data Science Lab at the University of Liverpool and the International Organization for Migration's (IOM) Displacement Tracking Matrix. This work has been developed in close partnership with operational agencies, ensuring its relevance to real-world needs and its potential to inform evidence-based decision-making.

The University of Liverpool and IOM Displacement Tracking Matrix recognize the value of data triangulation in enabling partners, including local authorities, to enhance coordination and programming through robust, evidence-based strategies. We welcome constructive conversations on how this approach can be further refined and applied, and we look forward to collaborating with stakeholders to promote sustainable, rights-based solutions for displaced populations worldwide.

**Professor Tim Jones**
**Vice-Chancellor**
**University of Liverpool**

**Laura Nistri**
**Global Displacement Tracking Matrix (DTM) Coordinator**
**IOM**

# CREDITS AND ACKNOWLEDGEMENTS

This report is the result of a collaborative effort between the Geographic Data Science Lab at the University of Liverpool and the International Organization for Migration's Displacement Tracking Matrix (DTM). The authors wish to express their sincere gratitude to colleagues at IOM Ukraine's Data and Analytics Unit and IOM DTM London. Special thanks to Douglas Leasure, Andrea Aparicio Castro and Edith Darin of the University of Oxford for their valuable contributions to the project and its outcomes.

Additionally, sincere thanks are extended to those who participated in the project's three workshops and hackathon, the names of which are listed below. The content of this report would not have been possible without the generous contribution of participants; their comments and insights aided greatly in shaping the findings and conclusions of this report. Further thanks are also extended to organizations that helped provide data and technical support during the project's hackathon, namely Snowflake and Direct Relief, both of whom the event and its success would not have been possible without. We gratefully acknowledge Ellen Van de Weghe (IOM DTM Pakistan) and Fawad Qureshi (Snowflake) for their valuable contributions as judges during the hackathon.

Workshop attendees:

Andrea Aparicio
Mohamed Bakr
Luong Bang Tran
Adam Bekele
Jos Berens
Dominik Bursy
Carmen Cabrera
Dan Caspersz
Alexander Chilton
Flora Chu
Franziska Clevers
Laura Coskun
Rachel Cribbin
Edith Darin
Adham Enaya
Marianthe Evangelidis
Gabriele Filomena
Emma Goatman
Joseph Goodall
Ali Guenduez
Shannon Hayes
Tony Hoad
Linh Hoang Thuy
Kate Hodkinson
Mike Johnson
Dirk Jung
Damien Jusselme
Varun Khandelwal
Cara Kielwein
Doug Leasure
Nando Lewis
Lisa Lim Ah Ken
William Lumala
Alice Marshall
Alex McCarthy
Brian McDonald
Andrea Nasuto
Euan Newlands
Abdul Samad Omari
Oluwatosin Orenaike
Benjamin Pfau
Elena Philipova
Joshua Phillippo-Holmes
Fawad Qureshi
Lorenzo Sileci
Joseph Slowey
Yaroslav Smirnov
Ellen Van de Weghe
Vivianne Van der Vorst
Ana Varela Varela
Belinda Volans
Huan Wang
Nick Ward
Scott White
Michael Zihanzu

This report was made possible through the support of the Policy Support Fund at the University of Liverpool, funded by Research England. The research aligns with the strategic objectives of UK Research and Innovation (UKRI) by enhancing social and economic resilience through the development of tools that help communities and policymakers more effectively manage displacement challenges. It also contributes to national security and risk awareness by improving the capacity to anticipate and respond to displacement crises. Furthermore, the project promotes the rights, dignity, and long-term well-being of displaced populations by advancing data-driven strategies that support sustainable integration or return.

# LIST OF FIGURES AND TABLES

# CHAPTER 1

## DATA TRIANGULATION IN CRISIS RESPONSE: BRIDGING TRADITIONAL AND DIGITAL SOURCES

Internal displacement remains a critical global issue. An unprecedented 83.4 million people were living in internal displacement at the end of 2024, according to the newly released Global Report on Internal Displacement 2025 (IDMC, 2025a). This scale of displacement underscores the urgent need for innovative, data-driven approaches to track and understand population movements. Traditional data systems provide vital information for those responding to humanitarian crises. However, as human mobility patterns become increasingly complex, the need for reliable, timely and spatially detailed data to inform the development of policy and humanitarian response is becoming increasingly acute. Traditional data streams are often not well-equipped to meet these needs (IOM, 2018b).

This report, a collaboration between the Geographic Data Science Lab at the University of Liverpool and the International Organization for Migration's (IOM) Displacement Tracking Matrix (DTM), examines how traditional data sources can be effectively integrated with emerging digital trace data – such as mobile phone GPS and social media activity – to enhance the monitoring of displacement in humanitarian settings. By leveraging these diverse data streams, the report demonstrates how triangulation can improve the precision and reliability of displacement estimates. This approach is informed by lessons learned from recent crises, particularly the escalation of the war in Ukraine and the 2022 floods in Pakistan. Building on the innovative application of digital trace data and the development of robust data infrastructure in these contexts, the report outlines a scalable framework for triangulating data across a broader range of crises, including those triggered by natural hazards and public health emergencies. By leveraging real-time, high-resolution data sources we aim to create a more responsive, scalable and accurate system for understanding displacement, allocating resources and evaluating interventions. This report presents the findings of a structured pilot effort designed to test this approach in two defined contexts, leading to a set of key conclusions that inform the future design of data triangulation systems for humanitarian action – followed by detailed technical insights and recommendations.

The report is split into three sections. First, we examine the current data landscape around displacement in conflict and disaster contexts. We provide an overview of the gaps in traditional data streams available during displacement events, and the limits these put on effective humanitarian action. Traditional data streams refer to established sources of information used in humanitarian contexts, such as surveys, administrative records and key informant interviews. We then explore the potential for digital nontraditional data to complement traditional data, provide additional insight and enhance efforts to respond to conflict and disaster events. Digital nontraditional data, or digital trace data, refers to information generated passively or actively through digital platforms and devices, such as mobile phone GPS signals, social media activity, satellite imagery and online transactions.

In its second section, the report focuses on the war in Ukraine, which escalated in February 2022. Statistical indicators derived from digital trace data sources – including displacement rates and return rates – are presented. These indicators are benchmarked against IOM data on Ukraine to assess their validity and enhance confidence in the use of digital trace data for monitoring displacement dynamics. This section provides links to technical documentation for all datasets, as well as publicly available code, to ensure the project's methodology is transparent and replicable in other conflict and disaster contexts. The third section of the report builds on the innovative data triangulation methods used in Ukraine to quantify conflict-related displacement, applying them to the 2022 Pakistan floods and showcasing how these approaches were tested during a hackathon to evaluate their broader applicability across crises emergencies. The report concludes by summarizing the value that digital nontraditional data can add and its significant benefit to humanitarian response in displacement events.

This work bridges a critical gap in the humanitarian sector by advancing innovative, data-driven approaches to displacement analysis. Its impact will extend to strengthening humanitarian response, improving policy formulation and fostering resilient, rights-based solutions for displaced populations worldwide.

**Table 1: Traditional data and digital nontraditional data definitions**

| Traditional data | Traditional data streams refer to established sources of information used in humanitarian contexts, such as surveys, administrative records and key informant interviews. |
|---|---|
| Digital nontraditional data | Digital nontraditional data, or digital trace data, refers to information generated passively or actively through digital platforms and devices, such as mobile phone GPS signals, social media activity, satellite imagery and online transactions. |

# CHAPTER 2

## KEY LESSONS ON TRIANGULATING DATA FOR DISPLACEMENT ESTIMATES IN DISASTER CONTEXTS

**This section provides an overview of the potential for triangulating multiple data sources to estimate population displacement in disaster settings, whether caused by conflict or climate-induced natural hazards. The insights presented here are informed by a series of expert workshops held in early 2025, which brought together stakeholders from government, academia, international organizations and the private sector.**

The workshops included:

**Shaping a Policy-driven Framework for Displacement Estimates**
13 February 2025, Online

**Engaging Government Stakeholders for Practical Insights**
25 March 2025, London

**Presenting the Framework and Exploring Broader Contexts**
21–22 May 2025, Berlin

Participants included representatives from the International Organization for Migration's (IOM) Displacement Tracking Matrix (DTM), the IOM Global Migration Data Analysis Centre (GMDAC), the IOM Ukraine Data and Analytics Unit, the UK Foreign, Commonwealth & Development Office (FCDO), the Centre for Humanitarian data at United Nations Office for the Coordination of Humanitarian Affairs (OCHA), the London Stock Exchange Group (LSEG), Snowflake and several non-governmental organizations (NGOs), including Direct Relief, Aid Ukraine and Operations for Change. Academic institutions represented included the University of Oxford, the London School of Economics (LSE), the University of Exeter and the University of Liverpool (UoL).

The workshops highlighted the critical role of multi-source data triangulation in improving the accuracy and responsiveness of displacement estimates. This approach is particularly valuable in rapidly evolving crises, where timely, high-resolution information can significantly enhance the targeting and effectiveness of humanitarian interventions. Digital nontraditional data sources – such as mobile phone records, social media and satellite imagery – offer high spatial and temporal granularity, while traditional survey methods contribute essential contextual understanding and validation. When used in combination, these traditional and nontraditional data sources provide a more comprehensive and reliable foundation for informed decision-making in humanitarian response planning.

## 2.1 DISPLACEMENT DATA SYSTEMS: CURRENT MODELS AND THEIR LIMITATIONS

Internal displacement has reached a record high. According to the Internal Displacement Monitoring Centre (IDMC), 83.4 million people were living in internal displacement worldwide at the end of 2024 – the highest number ever recorded and more than double the figure from six years earlier (IDMC, 2025a). The scale of internal displacement has severe consequences, pushing millions into precarious living conditions and undermining access to essential services such as health care, education and livelihoods. It also poses a significant barrier to achieving sustainable development. As the crisis of internal displacement has grown, the need for high-quality data has become more important than ever. Data systems providing accurate and timely information are crucial for quantifying the scale of displacement, the allocation of resources in disaster response, the monitoring of crisis events over time and for the evaluation of the effectiveness of interventions (see Figure 1).

**Figure 1. An overview of displacement data systems, data-collection methods and limitations**

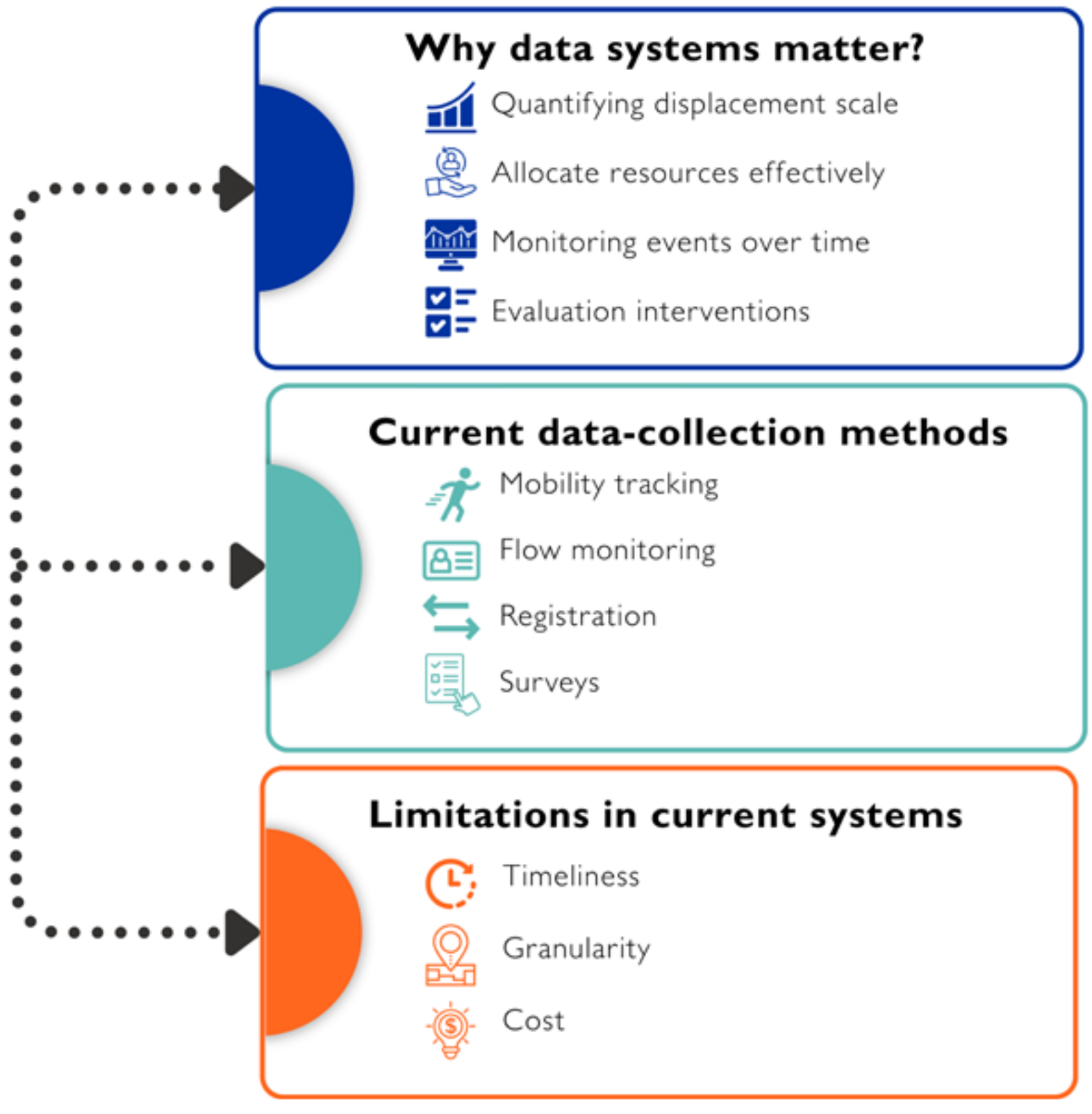

*Source:* Elaborated by the authors. "Current data-collection methods" sourced from the International Organization for Migration's Displacement Tracking Matrix (DTM) available at: https://dtm.iom.int/about/infosheets.

Table 2 presents the five primary datasets discussed in this report and highlights their respective strengths and trade-offs. These include both traditional and nontraditional data sources used to estimate internally displaced persons (IDPs) in humanitarian contexts. IOM's General Population Surveys (IOM RDD) in Ukraine are repeated cross-sectional sample surveys designed to provide reliable, ground-level insights into displacement and mobility trends. Each survey round collects responses from an independent sample using a consistent set of questions, with data gathered through Random Digit Dialling (RDD) and Computer-Assisted Telephone Interviews (CATI) (IOM, 2024a).

IOM DTM's Community Needs Identification (CNI PK) was established following the widespread flooding in Pakistan in 2022. It provides information on the number of displaced persons, along with the multisectoral needs of communities, at the settlement and village level (IOM, 2023). CNI PK is implemented through a key informant survey, in which IOM enumerators interview community leaders or representative groups to gather data and estimate displacement figures. The assessment was conducted in multiple rounds, targeting specific settlements and villages during each phase.

In contrast, the digital trace data sources presented in Table 2 – namely GPS Phone Data (GPS MD), Meta's Marketing Platform API (Meta MAPI) and Meta's Data for Good products (Meta DfG) – are all derived from private sector platforms. These datasets capture either geo-located observations of mobile devices at the coordinate level (GPS MD) or aggregated information on social media users across spatial units such as grid cells or administrative boundaries.

## CURRENT MODELS FOR DATA COLLECTION

Existing data systems are wide-ranging and rely on a combination of collection techniques to produce comprehensive insights (IOM, 2023b) (see Figure 1). In this report we consider both displacement estimates derived from key informant interviews (CNI PK) and a representative, repeated cross-sectional sample survey IOM RDD) (Table 2).

Key informant interviews remain one of the most-commonly used approaches for data collection in disaster contexts, as demonstrated during the 2022 floods in Pakistan (SNI PK). In this method, trained enumerators gather information from local officials or community leaders, enabling estimates of displaced populations and the identification of multisectoral community needs. This approach allows humanitarian organizations to collect data in a relatively time- and resource-efficient manner, often generating timely operational insights. However, each phase of data collection often targets only selected settlements and villages, which limits the generalizability of findings to affected areas. The method also requires significant human resources and can encounter access constraints, leading to delays between the onset of the disaster, data collection and subsequent reporting.

Data-collection methodologies are generally tailored to specific country or disaster contexts. Where full access to displaced populations is possible, humanitarian organizations may implement or utilize detailed registration techniques to capture survey-like data through interviews with individuals or households (Carletto et al., 2022; Kilic et al., 2017). For example, IOM DTM displays data collected by the Ukrainian Ministry of Social

### Table 2. Data comparison of primary datasets discussed

| Acronym | Data | Provider | Geographic unit | Frequency | Type | Unit | Access | Processing |
|---|---|---|---|---|---|---|---|---|
| IOM RDD[a] | General Population Survey Ukraine | IOM | Admin 1 (oblast) Admin 2 (raion) | Initially every 2 months; quarterly from Round 13 onward | Sample survey | Respondent | High cost | Manpower |
| CNI PK[b] | Community Needs Identification Pakistan | IOM | Settlement/ Village (aggregated up to Admin 1 and Admin 2) | Depends on funding and government approvals | Key informant | Settlement/ Village | High cost with some access issues | Manpower |
| GPS MD[c] | GPS Phone Data | Private company | GPS | Hourly | Multiple Apps | Mobile device | High cost | Computational |
| Meta MAPI[d] | Meta's Platform Marketing API | Private company | Area unit (e.g. Admin 2) | Daily | Single App | User (age and sex) | Free | Computational |
| Meta DfG[e] | Meta Data for Good | Private company | Area unit (grid or admin-level) | Subdaily | Single App | User | Free | Computational |

*Notes:* a. The IOM RDD General Population Survey in Ukraine is a repeated cross-sectional study that uses a consistent questionnaire administered to new respondents each round. In its first 12 rounds, a random digit dial (RDD) method was used to survey 2,000 adults (18+) living in government-controlled areas of Ukraine, yielding a ±2.0 per cent margin of error at a 95 per cent confidence level, with results extrapolated nationally and by macroregion. From round 13, the sampling approach was revised to improve oblast-level estimates for IDPs, returnees and non-displaced populations, initiating a second research cycle (IOM, 2024a). The sample population consists of the general population of Ukraine aged 18 and over who are currently in the country at the time of the surveys conducted.
b. CNI PK refers to the Community Needs Identification after the widespread flooding in 2022 in Pakistan and provides information on the number of displaced persons and the multisectoral needs of communities at the settlement/village level.
c. GPS MD refers to anonymized geolocation information collected from mobile devices, capturing users' movements over time.
d. Meta MAPI provides aggregated, anonymized audience estimates based on user demographics and location, offering a valuable proxy for tracking population distribution and mobility in near real time (Leasure et al., 2023).
e. Meta DfG refers to two related datasets provided by the platform: Facebook Population During Crisis and Facebook Movement During Crisis; these provide stock and flow data of Facebook users, respectively and are made available by Meta in the aftermath of a crisis.

Policy on the total number of registered IDPs on an online dashboard (IOM, 2025a). Due to cost, further information on IDPs is usually drawn from samples of the displaced population. IOM DTM's own data collection in Ukraine has been based around the General Population Survey (IOM RDD). This was rolled out in the aftermath of the full-scale invasion by the Russian Federation in 2022 and utilized a Random Digit Dialling (RDD), leveraging the country's high rate of mobile phone penetration (IOM, 2022a).[1] This method initially enabled the collection of nationally and macroregionally representative data on displacement and return, while also capturing contextual information to guide policy and response. The sampling was later expanded to provide oblast-level estimates for more granular analysis. Despite their robustness, such surveys have limitations – including lower spatial resolution (typically at administrative level 1 or 2) and high operational costs (IOM, 2024a). Additionally, while conducted quarterly, their frequency cannot match the near real-time insights offered by digital trace data.

[1] The methodology underwent several revisions across rounds. In 2022 (Rounds 1–12), data collection followed a single-phase design with 2,000 respondents per round. In 2023 (Rounds 13–16), a multiphase, two-stage design was introduced, expanding the sample to 20,000 respondents in Phase 1, with stratified probability proportional to size (PPS) sampling at the second stage, based on oblast and displacement status. In 2024 (Rounds 17–20), the approach was further expanded to 40,000 respondents in Phase 1, with a revised Phase 2 sampling strategy to ensure evenly disaggregated sample targets across strata, achieving an overall target of 4,800 respondents (IOM, 2024a).

In other contexts, such as Somalia, populations are highly mobile due to a combination of recurrent conflict and environmental pressures, including frequent flooding. To better understand the scale and dynamics of population movements in such settings, flow monitoring techniques are often applied, although these focus on cross-border flows (IOM, 2024c). The IOM has established a Flow Monitoring Registry (FMR) to track the number of individuals passing through key transit locations. This system relies on brief interviews with key informants and is implemented following baseline assessments to determine the most appropriate flow monitoring points. In other settings, such as Yemen, IOM's DTM also conducts flow monitoring surveys to gather detailed information on the demographics, characteristics, origins and motivations of cross-border migrants (IOM, 2024b).

## LIMITATIONS OF CURRENT DATA SYSTEMS

These IOM-deployed data gathering technologies provide methodologically robust estimates of displaced people and crucial demographic and community needs information. In fact, traditional data streams are of paramount importance in aiding humanitarian response during displacement events, and for the design of policies to mitigate the effect of disasters. However, they have limitations.

The most significant challenge faced by current data systems is their inability to provide estimates of population movements in both a timely enough manner and at a granular level of spatial detail (Green et al., 2023). Traditional data systems can suffer from a lag between data collection and publication, or may experience access issues, preventing primary data from being collected for an extended period. In cases where data are made available very rapidly, they are often not able to provide a spatially detailed picture of displacement. The CNI PK was completed a few months after the initial flood event, and findings could not be generalized as the data were collected from a selected sample of villages in Pakistan (IOM, 2023). The IOM RDD, on the other hand, provided estimates of IDPs within less than a month of the Russian Federation's full-scale invasion; however, the information was available at a macroregion level and the insights obtained could provide limited evidence to support resources allocation efforts at a localized level (IOM, 2022).

Traditional data systems, while timely in a conventional sense, are not real time. This gap poses challenges for those responding to displacement crises. Without rapid estimates of displaced populations, humanitarian organizations and governments struggle to quantify response needs, while limited spatial detail restricts the efficient allocation of already scarce resources across affected areas. Moreover, traditional survey-based methods that require on-the-ground collection of data (such as in the case of key informant surveys) are often difficult to conduct.

Enumerators can face security and operational challenges that limit their ability to reach displaced populations and communities. In regions experiencing disaster and displacement events, these challenges are heightened and primary data collection may even prove too difficult to conduct at all.

Participants from humanitarian and policymaking organizations raised concerns during the workshops about data limitations that had constrained the effectiveness of disaster response. In Indonesia, for example, as described by workshop attendants, slow data collection and limited spatial detail on displaced populations left responders uncertain about which communities needed urgent support (IOM, 2018a). This restricted effective resource allocation and resulted in poor documentation of displacement patterns, undermining the development of evidence-based policies for future crises.

Digital nontraditional data sources offer higher spatial and temporal granularity. GPS MD, for example, can capture hourly device movement at the coordinate level, while Meta MAPI and Meta DfG datasets provide user demographic insights at daily or even subdaily intervals. However, each digital source has its own limitations, such as varying app penetration rates, potential biases in user demographics and computational demands (Rowe, 2023; Hodkinson, 2025). Furthermore, digital nontraditional data sources require expert technical knowledge to turn raw messy and unstructured data into insightful and actionable information (Arribas-Bel et al., 2021). Table 3 provides brief descriptions of the digital trace data sources examined in this report and summarizes their key strengths and limitations. It highlights the value of each source for tracking population movements, estimating population presence and informing humanitarian response.

### HUMANITARIAN DATA UNDER THREAT

The humanitarian and development sectors have faced significant funding gaps for many years, affecting both the collection of data and delivery of data. Recent funding cuts to global humanitarian operations have exacerbated this trend, raising serious concerns about the sustainability of essential data systems. As noted in The New Humanitarian (Stoddard et al., 2025), "the data streams that underpin humanitarian response are about to collapse." This concern was echoed across discussions during the project's workshops, where stakeholders highlighted how funding reductions have threatened the availability of critical data for informing disaster response and policymaking.

This context presents significant challenges. Reduced access to timely, reliable data makes it more difficult to anticipate needs, allocate resources and coordinate responses – especially in complex emergencies. While digital nontraditional data cannot replace the full depth and contextual insight provided by traditional data collection, such data can help fill critical information gaps, particularly when conventional data systems are weakened or underfunded.

Some digital nontraditional data sources – such as social media and platform-based population estimates – are available at little or no cost. Others, such as GPS phone data, can be more expensive and may require substantial investment or data-sharing partnerships for access. Furthermore, the integration of these data requires careful consideration of data privacy and ethics, particularly in cases where standards between the organizations sharing these data differ. Regardless, the value of digital nontraditional data has become apparent. They offer a scalable and flexible alternative to complement or even fill gaps when traditional data streams are disrupted. Currently, there is an urgent need for greater coordination across the humanitarian, governmental, academic and private sectors, to tackle the ongoing data infrastructure crisis and coordinate data efforts integrating nontraditional data as part of the humanitarian data ecosystem. Collaborative systems for data-sharing and insight generation can help maximize the value of available data while reducing the financial burden on any single organization. Participants across all three workshops expressed strong support for the development of such frameworks, which would promote more efficient use of data, foster cross-sector collaboration and improve the resilience of humanitarian information systems.

## 2.2 TURNING DIGITAL TRACE DATA INTO OPPORTUNITIES

### The potential of digital trace data in crisis response

**Figure 2. High resolution, large-scale coverage and near real-time availability**

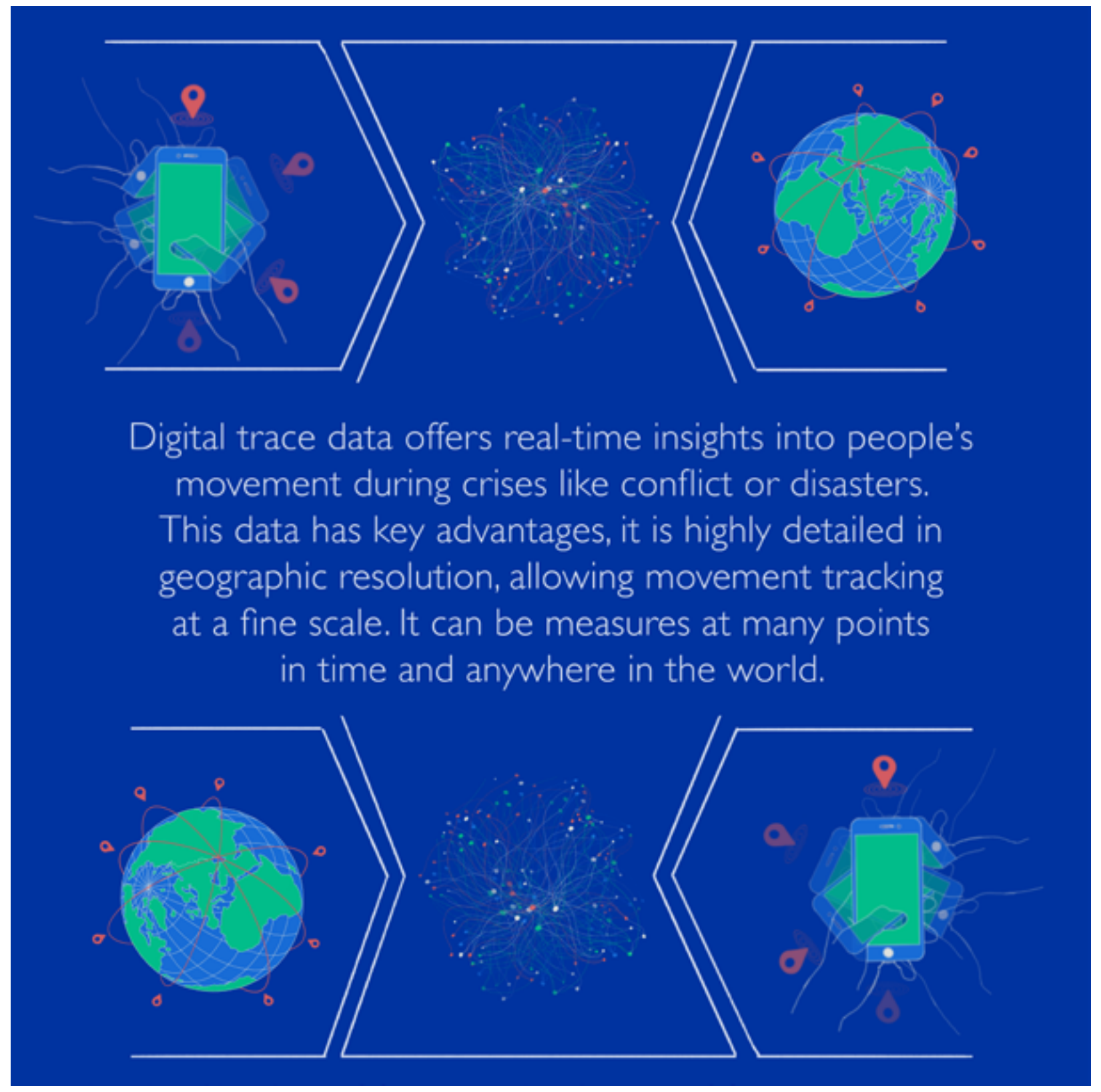

*Source*: Elaborated by authors. Designed by Lea Riggi.

**Figure 3. Hard-to-reach location access**

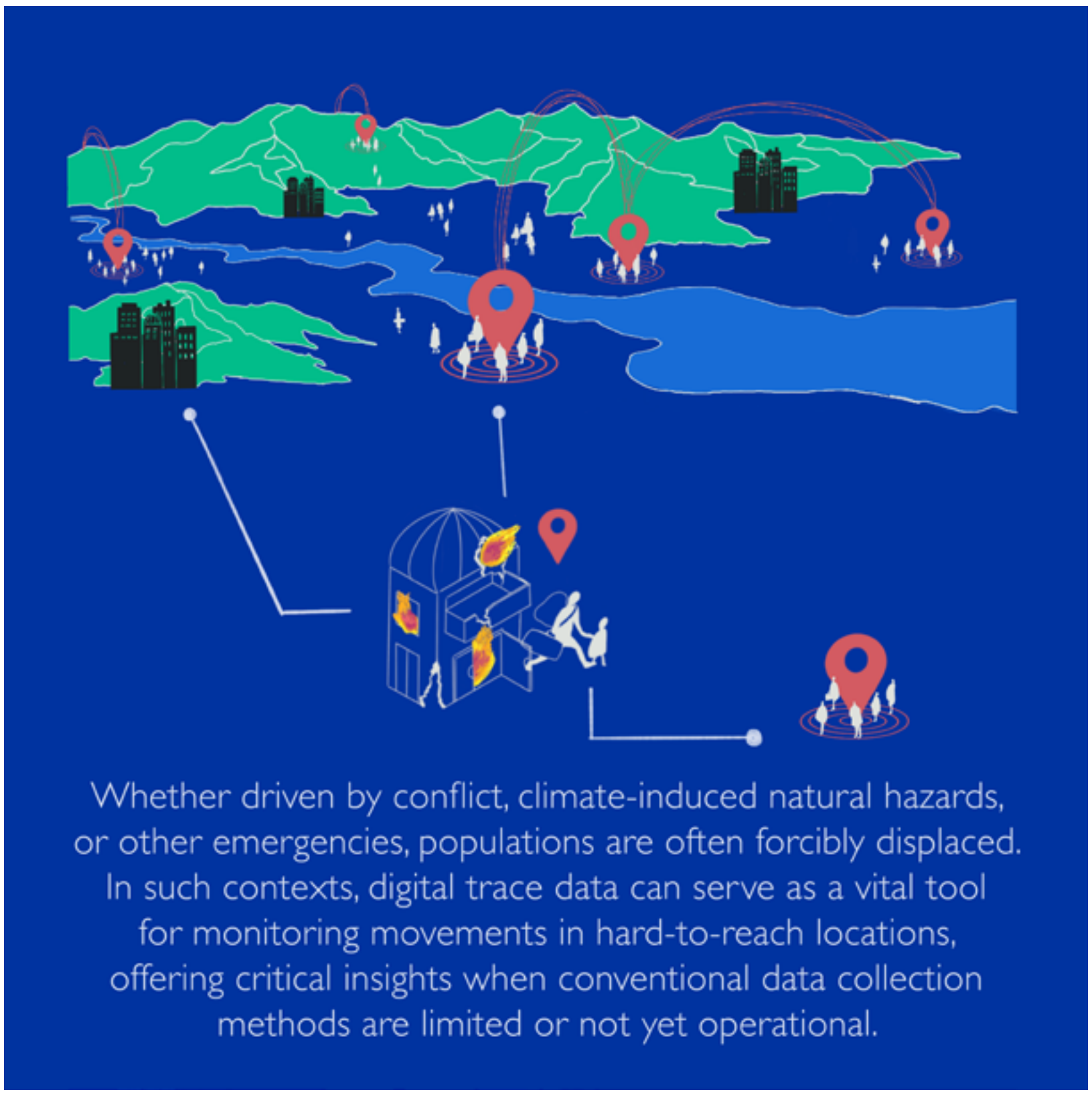

*Source*: Elaborated by authors. Designed by Lea Riggi.

Advances in digital technology and their growing global adoption present a significant opportunity to address existing data gaps in humanitarian response. Digital trace data can offer timely, high-resolution information, enabling policymakers to gain localized insights as soon as a crisis unfolds.

### High resolution, large-scale coverage and near real-time availability

This data draws from a broad range of sources – including mobile phone data, social media and satellite imagery. It offers several key advantages (Rowe, 2023). It provides high geographic resolution, enabling the estimation of population movements at a fine scale; temporally frequent observations offering near real-time quantification of movements; and large space geographical and population coverage potentially affording the estimation of population movement at any location across the globe (Figure 2).

### Hard-to-reach location access

During crises – whether driven by conflict, climate-induced natural hazard hazards or other emergencies – populations are often forcibly displaced. In such contexts, digital trace data can serve as a vital tool for monitoring movements in hard-to-reach locations, offering critical insights when conventional data-collection methods are limited or unavailable (Figure 3).

### Actionable insights through geographic data science

Spatial data analysis and geographic data science techniques are needed to extract meaningful insights from digital trace data. These methods include data visualization, statistical modelling, and geospatial scripting and analysis. For instance, as detailed in Section 2 of this report, GPS-based mobile phone data can be transformed from raw counts of devices into reliable estimates of population presence and movement (Iradukunda et al., 2025). This process illustrates how advanced analytical tools can convert complex digital trace data into actionable information for humanitarian response (Figure 4).

**Figure 4. Actionable insights through geographic data science**

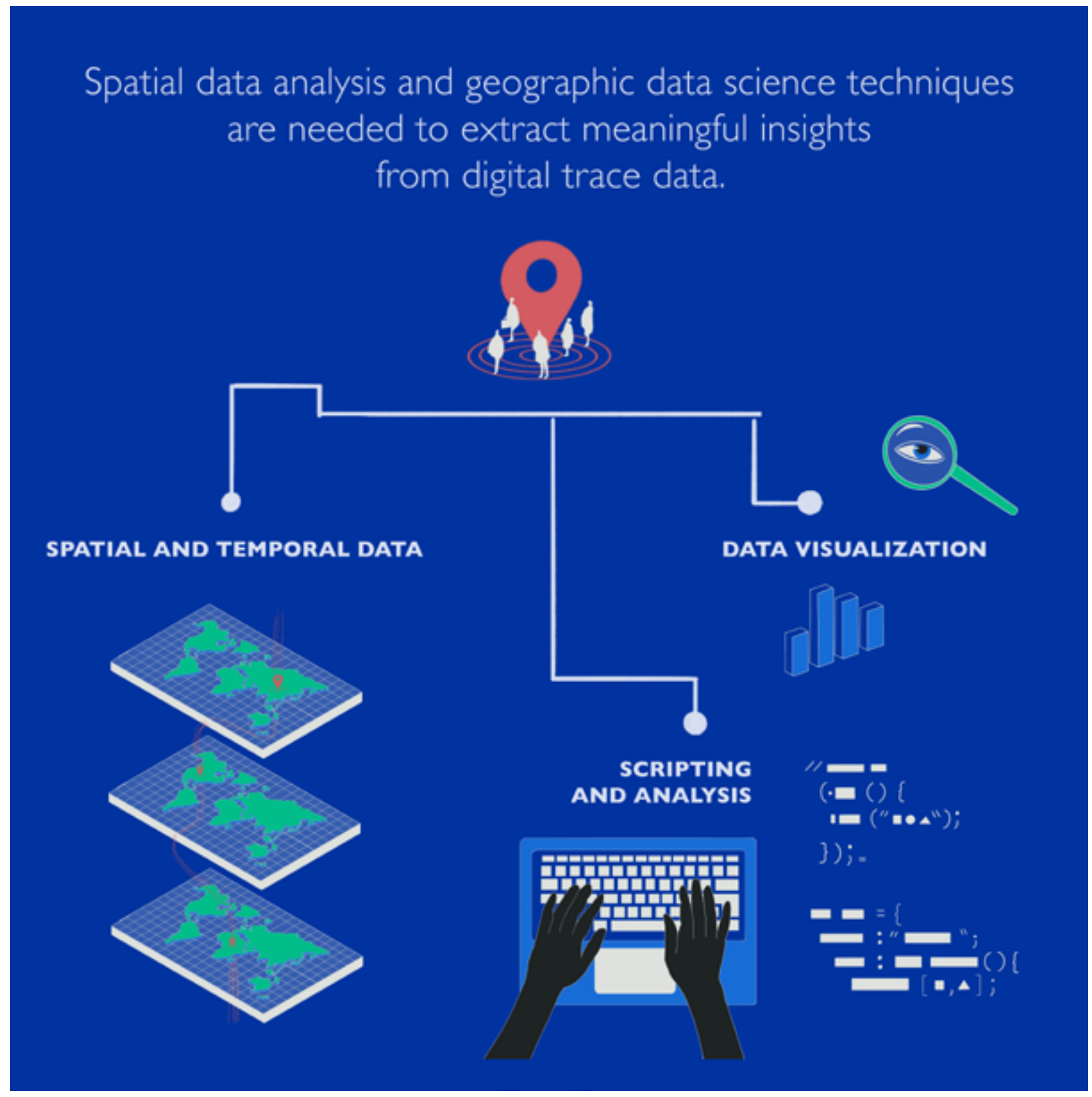

*Source*: Elaborated by authors, Designed by Lea Riggi.

**Table 3. Comparison of digital nontraditional data sources for crisis response and displacement analysis**

| | Description | Strengths | Limitations |
|---|---|---|---|
| **GPS (Global Positioning System) Phone Data** | • Anonymized geolocation information collected from mobile devices showing users locations and movements | • Offers real-time, high-frequency data, often at an hourly level<br>• Captures fine-grained movement patterns at the coordinate level<br>• Allows early insight into displacement trends from the onset of a crisis<br>• Useful for estimating routes and identifying arrival areas for displaced populations | • App users do not provide explicit informed consent, reliance solely on general terms and conditions<br>• Lacks demographic or contextual detail<br>• Requires agreements with telecom providers for access<br>• Limited usefulness in areas with low mobile phone penetration |
| **Facebook's Population During Crisis** | • Stock data provided by Meta's Data for Good platform showing the number of Facebook users within a specific area<br>• More details can be found here: https://dataforgood.facebook.com/dfg/tools/facebook-population-maps | • Provides data on population presence and stocks across administrative areas and small geographic tiles (e.g. $800m^2$)<br>• Includes baseline data for comparison with pre-crisis period<br>• Freely accessible to humanitarian actors through Meta's Data for Good platform<br>• High temporal resolution enables tracking changes in near real time | • App users do not provide explicit informed consent; data use is based solely on general terms and conditions; Meta restricts data availability through minimum user threshold requirements<br>• Covers only Facebook users with location settings enabled, with platform usage varying by region<br>• May not represent full population demographics<br>• Available only during recognized crises<br>• Data availability limited to 90 days post-event<br>• Limited ability to customize or adjust data requests |
| **Facebook's Movement During Crisis** | • Flow data provided by Meta's Data for Good platform showing movement of Facebook users between origin and destinations pairs<br>• More details can be found here: https://dataforgood.facebook.com/dfg/tools/movement-maps | • Provides data on population movement and flows across administrative areas and small geographic tiles (e.g. $800m^2$)<br>• Includes baseline data for comparison over with pre-crisis period<br>• Freely accessible to humanitarian actors through Meta's Data for Good platform<br>• High temporal resolution enables tracking changes in near real time | • Meta restricts data availability through minimum user threshold requirements<br>• Covers only Facebook users with location settings enabled, with platform usage varying by region<br>• May not represent full population demographics<br>• Available only during recognized crises<br>• Data availability limited to 90 days post-event<br>• Limited ability to customize or adjust data requests |
| **Meta Marketing Platform API** | • Stock data sourced from Meta's marketing API showing separately the number of Facebook and Instagram users by age and gender within a specific area | • Enables estimates of user populations in small areas through advertiser tools<br>• Offers basic demographic data (age, gender)<br>• Can detect sudden changes in populations stocks<br>• Accessible before, during and after crises – but the pipeline to retrieve data needs to be set up before or when the crises begins as historical data cannot be queried from the API | • Limited to Meta platform users, with coverage varying widely<br>• Requires technical skills to query and interpret data<br>• Platform terms and accessibility change over time<br>• Estimates are made available by the NowPop group following the guidelines from the Information Sharing Protocol for Ukraine set by UNOCHA |

## 2.3 FROM INSIGHT TO ACTION: RECOMMENDATIONS FOR INTEGRATING DIGITAL TRACE DATA IN HUMANITARIAN RESPONSE

Based on the in-depth discussions with stakeholders across the three workshops, the recommendations illustrated in Figure 2 have been devised. These are essential in ensuring digital nontraditional data are more widely adopted in humanitarian response and are successfully integrated into decision-making processes.

**Figure 5. Recommendations for integrating digital trace data in humanitarian response**

**Standardize data definitions**
Maintain consistency in key terms to ensure comparability across data sources.

**Rapid and actionable data products**
Deliver timely insights - within 2 weeks to 1 month post-disaster - to inform early decision-making before traditional data becomes available.

**Share data responsibly**
Develop systems for responsible dissemination and actionable, policy-relevant data-sharing.

**Foster trust**
Engage governments, the private sector, academia and local communities to build confidence and encourage collaboration.

**Prioritize ethics**
Ensure responsible use of data by addressing privacy, consent and data protection.

**Ethical data engineering**
Develop systems for efficient and safe data processing

**Acknowledging limitations and ensuring transparency**
Clearly communicate data gaps, biases, assumptions and technical constraints to support responsible and informed use.

**Clarify methods**
Ensure methodologies are well-documented, transparent and replicable to support credibility and understanding.

**Promote data generalizability**
Strive for methods and estimates that can be used beyond a specific case-study to broader contexts and applicability.

**Integrating with existing systems**
Integrate digital data with established humanitarian data streams for maximum impact.

*Source*: Elaborated by authors.

### Effective use

#### 1. Standardized data definitions

To enable comparability across data sources, it is essential to maintain consistency in key terms and definitions – particularly for core concepts like "internally displaced persons (IDPs)" and "returns". Where definitions differ, these distinctions should be clearly documented and made transparent to end users. When transforming digital trace data for humanitarian use, alignment with existing data systems is critical. This includes using population group classifications and geographic units familiar to humanitarian actors. For example, OCHA's use of "PCodes" to define subnational geographies is a standard within the sector, though they are rarely applied in other domains. Aligning digital trace data outputs with these established geographic conventions ensures compatibility and facilitates integration with traditional data streams. Without such alignment, meaningful comparison and joint analysis across sources become significantly more difficult.

### 2. Integrating nontraditional data into existing data systems

As described previously, digital nontraditional data can offer real time, granular insights before traditional data-collection systems are operational, enabling faster decision-making at the onset of displacement events. In today's climate of funding cuts, these novel data sources can also help fill critical information gaps in resource-constrained settings. To fully harness their potential, digital trace data must be meaningfully integrated into traditional humanitarian data systems. OCHA's Humanitarian Data Futures project offers a useful model for adoption, based on a four-part cycle: (1) Technologies – technologies enable improvements in (2) Infrastructure and systems – that allow them to be deployed into (3) Applications and use cases – these are then used to create value according to (3) Social norms and values – the value created supports stakeholder trust and investment in future technologies, restarting the cycle (Hodkinson, 2025; Sharpe et al., 2016). While this cycle can support the normalization of digital nontraditional data use, adoption is rarely linear. Many innovations fail to take root unless they clearly address existing data needs and align with current systems. Ensuring that digital nontraditional data can effectively complement and connect with traditional sources, is key to their long-term adoption and impact.

### 3. Promote data generalizability

Innovative data systems often struggle to scale. To support broader adoption, methodologies should be designed with generalizability in mind – producing estimates and outputs that extend beyond a single case study to wider contexts and use cases. Satellite data provides a strong example: over time, it has evolved into a dependable, widely trusted resource across sectors, including humanitarian response (Caribou Space, 2022). A similar trajectory is needed for digital trace data. However, these sources are diverse – ranging from mobile phone data to social media signals – and there is no single standard format. To move from isolated use to broader integration, we must develop approaches that allow different types of digital trace data to feed into existing humanitarian data structures in a systematic and scalable way.

## Timeliness and usability

### 4. Rapid and actionable data products

A consistent theme across all three policy workshops was the urgent need for data to be made available as quickly as possible. In disaster settings, speed is critical. Policymakers and humanitarian actors often rely on data that is "good enough" – not perfect, but timely and sufficient to support early decisions and resource allocation.

Digital nontraditional data has clear advantages in this space. Unlike traditional data systems, which are often delayed due to funding, organizational or security constraints, digital nontraditional data can be available in near real time and does not require physical access to affected areas. This makes it especially valuable in the chaotic early stages of a crisis. Participants highlighted digital nontraditional data as a powerful complement to early warning systems, particularly in regions with limited humanitarian presence or baseline data. It offers early signals for action and, retrospectively, can help reconstruct displacement pattern – crucial for planning future responses and pre-positioning resources. However, participants stressed that it is not the raw data, but the data product – the cleaned, analysed and interpreted output (Arribas-Bel et al., 2021) – that must be made trustworthy, timely and usable to truly support decision-making.

**5. Share data responsibly**

Clear rules on what data can be shared, with whom, and under what conditions help protect the rights and privacy of affected populations. Data should be presented in formats that are accessible and actionable for policymakers and humanitarian actors – such as clearly labelled data sheets or visual summaries. For example, donors like the UK Foreign, Commonwealth & Development Office (FCDO) often rely on concise, digestible data products to support rapid decision-making. Sharing information in a readable, ethical and secure way not only enhances response effectiveness but also fosters trust and collaboration across sectors.

## Methodological rigour and transparency

**6. Clear, transparent and well-documented methods**

Digital nontraditional data often arrives in raw, pre-processed formats that are not immediately usable by humanitarians or policymakers. To ensure trust and usability, the methods used to transform this data into actionable insights must be clearly documented, technically sound and easily understood. Processing pipelines should follow a transparent framework, with well-defined methodologies and accessible analysis scripts. Clear documentation – both technical and streamlined – is essential for enabling reproducibility and supporting collaboration across organizations.

Equally important is communicating these processes in a way that builds data literacy. Explaining how data was collected, processed and intended to be used – especially for non-expert audiences – will strengthen confidence in novel data sources and help integrate them more effectively into humanitarian decision-making.

**7. Ethical data engineering**

To maximize the utility of digital nontraditional data in humanitarian contexts, systems must be developed that enable efficient and ethical data processing. Secure, standardized pipelines – where raw inputs are transformed into anonymized, aggregated or insight-based outputs – help safeguard individual privacy while maintaining the value of the data. Clear governance frameworks are essential to guide decisions on how data is handled, ensuring that ethical principles are upheld throughout the process. Protocols for secure access, ethical review, and thorough documentation are critical for ensuring that sensitive information is processed with care. Identifying common ground between private sector ethics and humanitarian principles is also important, enabling the integration of ethical frameworks used by humanitarian actors with corporate practices to promote responsible data use and protect affected populations.

### 8. Acknowledging limitations and ensuring transparency

Effective use of digital nontraditional data requires transparency about its limitations. This includes communicating data gaps, biases and technical constraints (Cabrera and Rowe, 2025). Many sources – such as GPS phone or social media data – are shaped by uneven access to technology and have low rates of penetration in many countries. While this limits representativeness, these data can still provide value if carefully transformed to reflect broader populations and accompanied by clear explanations of their coverage and generalizability. By acknowledging limitations without undermining credibility, and by framing findings within their appropriate context, digital nontraditional data can support more responsible and effective decision-making.

Recognizing that insights from one case study may not apply universally is important. Displacement patterns and data quality vary across contexts, so findings must be interpreted within their specific setting. Privacy and confidentiality must be protected, particularly when handling disaggregated data. Responsible sharing requires clear governance frameworks, including ethical review, metadata publication and transparent documentation of methods. For instance, the granularity of IOM DTM data varies depending on whether it is publicly available or provided on request, with disaggregation and detail adjusted accordingly to balance openness with data protection. By framing limitations constructively and linking insights to original sources, digital nontraditional data can be integrated more effectively into policymaking and earn the trust of decision-makers.

## Ethics and trust

### 9. Prioritize ethics

Ensure responsible use of data by addressing privacy, consent, and data protection. In contrast to traditional survey methods, where informed consent is explicitly obtained from respondents, the use of digital trace data raises ethical challenges. App users typically agree only to the general terms and conditions of a service, without providing explicit consent for their data to be repurposed for further analysis. While companies such as Meta introduce safeguards – such as publishing only aggregated outputs or imposing minimum user thresholds – and GPS mobile phone data generally exclude direct demographic identifiers, these measures do not fully resolve the ethical concerns. Both cases highlight the importance of caution, as the boundary between acceptable use and potential misuse of such data remains a grey area, requiring robust governance frameworks and continuous ethical oversight.

Data-sharing is also a key consideration. Disaggregated digital nontraditional data can often contain information considered personally identifiable that would breach ethical guidelines if published or shared across organizations. This presents a clear ethical and reputational risk. As a result, clear processes should be established to produce anonymized, unidentifiable and aggregated data suitable for sharing, or to generate actionable insights from these data that can be safely provided to humanitarian organizations. Such a framework would protect the privacy of users' data but also enable a more efficient pipeline of insight generation in which only those conducting analysis need to work with raw data files.

### 10. Foster trust

Trust is a critical foundation for integrating digital nontraditional data into humanitarian action. Without it, adoption by governments, humanitarian agencies and policymakers will remain limited. While legal, ethical and institutional constraints present real challenges, they also offer an opportunity to build stronger, more inclusive partnerships among data providers, data warehouses, academia, humanitarian organizations and local communities. Strengthening these relationships can enhance transparency, ensure ethical data use, and ultimately increase the relevance and impact of digital nontraditional data in humanitarian contexts. Establishing collaborative frameworks – such as shared data pipelines, transparent methodologies and clear governance protocols – can facilitate integration and ensure responsible data use.

Initiatives such as the UK's Smart Data Services – Imago and GeoDS – demonstrate how data platforms can transform complex datasets into accessible, user-friendly products for a wider range of stakeholders beyond technical specialists, such as by converting raw GPS phone data into actionable insights. Similarly, models such as Flowminder demonstrate how academic non-profits can serve as trusted intermediaries for sensitive data-sharing, underpinned by legal frameworks and inter-agency agreements.

Participants also emphasized the value of developing harmonized data-sharing agreement templates, despite institutional barriers. There is a clear cross-sector appetite for a more unified approach, such as a global data-sharing framework, flexible data models to enable interoperability between agencies, shared institutional review boards, and national data libraries operating within open ecosystems but with strong access controls. The use of synthetic or anonymized data was also suggested to further reduce bureaucratic and legal hurdles, enabling earlier collaboration while protecting individual privacy.

Fostering this ecosystem of trust and transparency is essential – not only to strengthen institutional relationships, but to ensure the responsible, scalable and impactful use of digital nontraditional data for humanitarian and development goals.

# CHAPTER 3

## DIGITAL TRACE DATA IN ACTION: THE CASE OF UKRAINE

### 3.1 EVOLVING METHODOLOGIES IMPROVE ACCURACY OF DISPLACEMENT DATA

On 22 February 2024, the Russian Federation launched a full-scale invasion of Ukraine, marking a major escalation in the war (BBC, 2025). The invasion triggered widespread displacement and a humanitarian crisis. According to the United Nations High Commissioner for Refugees (UNHCR), as of 2025, there are 3.8 million internally displaced Ukrainians, 5.1 million refugees and 12.7 million Ukrainians in need of humanitarian assistance (IOM, 2025a; OCHA, 2025; UNHCR, 2025a). At the time of publication, the war in Ukraine remains ongoing.

#### Early displacement data is usually limited by conflict conditions

Estimating the scale of internal displacement during the onset of a crisis can be highly challenging. Widespread disruptions to infrastructure and essential services often hinder data collection, complicating efforts to monitor population movements in the early stages. Ongoing instability and limited access to affected areas further constrain the availability of accurate and timely information.

#### Evolving survey methodologies for enhanced representation in Ukraine

Recognizing the urgent need for reliable displacement data and leveraging Ukraine's high level of mobile phone penetration (SSSU, 2021), IOM Ukraine's Data and Analytics Unit launched a systematic, representative assessment of the Ukrainian population within ten days of the invasion. Preliminary findings were released shortly thereafter, with the first public report published in mid-March 2022 (IOM, 2024a). Using Computer-Assisted Telephone Interviewing (CATI) and Random Digit Dialling (RDD) methodologies, the initiative generated rapid insights into the size, distribution, mobility and immediate needs of the internally displaced population.[2] The resulting estimates were reviewed and validated by key humanitarian stakeholders (IOM, 2022b) and served as a foundation for timely humanitarian planning and enabling evidence-based advocacy throughout the country (OCHA, 2022).

Since 2022, IOM Ukraine has significantly strengthened its data-collection methods, producing more granular, timely and reliable insights to inform humanitarian operations

[2] The General Population Survey covered all oblasts of Ukraine and the city of Kyiv with active Ukraine-based cellular networks at the time of data collection, excluding the Autonomous Republic of Crimea and the city of Sevastopol, Ukraine, temporarily occupied by the Russian Federation and areas of Donetska, Luhanska, Khersonska and Zaporizka oblasts under the temporary military control of the Russian Federation where Ukrainian phone coverage was unavailable.

(IOM, 2024a). Between March 2022 and April 2025, 20 rounds of the General Population Survey (GPS) were conducted – initially on a bi-monthly basis (Rounds 1–12), and then quarterly from Round 13 onwards. For clarity, the GPS is referred to as IOM RDD throughout this report, as indicated in Table 2. In June 2024, the methodology was further refined to generate oblast-level representative data for more localized analysis. The IOM RDD sampling strategy was revised twice to enhance precision: beginning in June 2023 (Rounds 13–16), the sample size increased to 20,000 respondents per round, and from June 2024 (Round 17 onward), to 40,000. These adjustments aimed to improve estimates for IDPs, returnees and non-displaced populations at the oblast level, while also providing indicative data at the raion level.

For full methodological details – including the calculation of stocks of de facto IDPs and returnees, derived by extrapolating the shares identified in the survey to the baseline population – see Methodological Note – General Population Survey (IOM, 2024a).

## 3.2 DIGITAL TRACE DATA: ENHANCING CRISIS MONITORING IN REAL TIME

While the methodology and efforts in this case are to be commended, access to reliable data remains challenging in many conflict-affected areas for several reasons, including ongoing insecurity, restricted access due to frontline shifts, damaged infrastructure and the displacement of affected populations themselves. These obstacles often prevent survey data-collection methods from providing timely or accurate information, especially in the critical early stages of a crisis.

Digital nontraditional data offers a valuable alternative in such contexts, as it does not rely on active data-collection and can be analysed in near real time. This characteristic makes digital trace sources particularly useful immediately following a disaster or during the initial phase of conflict, when rapid and accurate information is crucial for prioritizing humanitarian needs and allocating resources effectively.

With this in mind, we triangulate three key data sources on displacement in Ukraine in 2022 – IOM RDD (survey data), GPS MD (mobile device GPS data), and META MAPI (social media-based displacement indicators) (Table 2) – to enhance the reliability and comprehensiveness of estimates, while also assessing the strengths and limitations of each source. This multisource triangulation is especially critical in fast-changing crises, where timely, high-resolution insights can dramatically improve the targeting and effectiveness of humanitarian interventions, balancing precision, cost and reliability.

*Nota bene*: For comprehensive technical details on the methodology of GPS MD and META MAPI – covering all stages from processing raw digital trace data to deriving population displacement estimates – please refer to the technical documentation provided in Iradukunda et al., 2025 and Leasure et al., 2023.

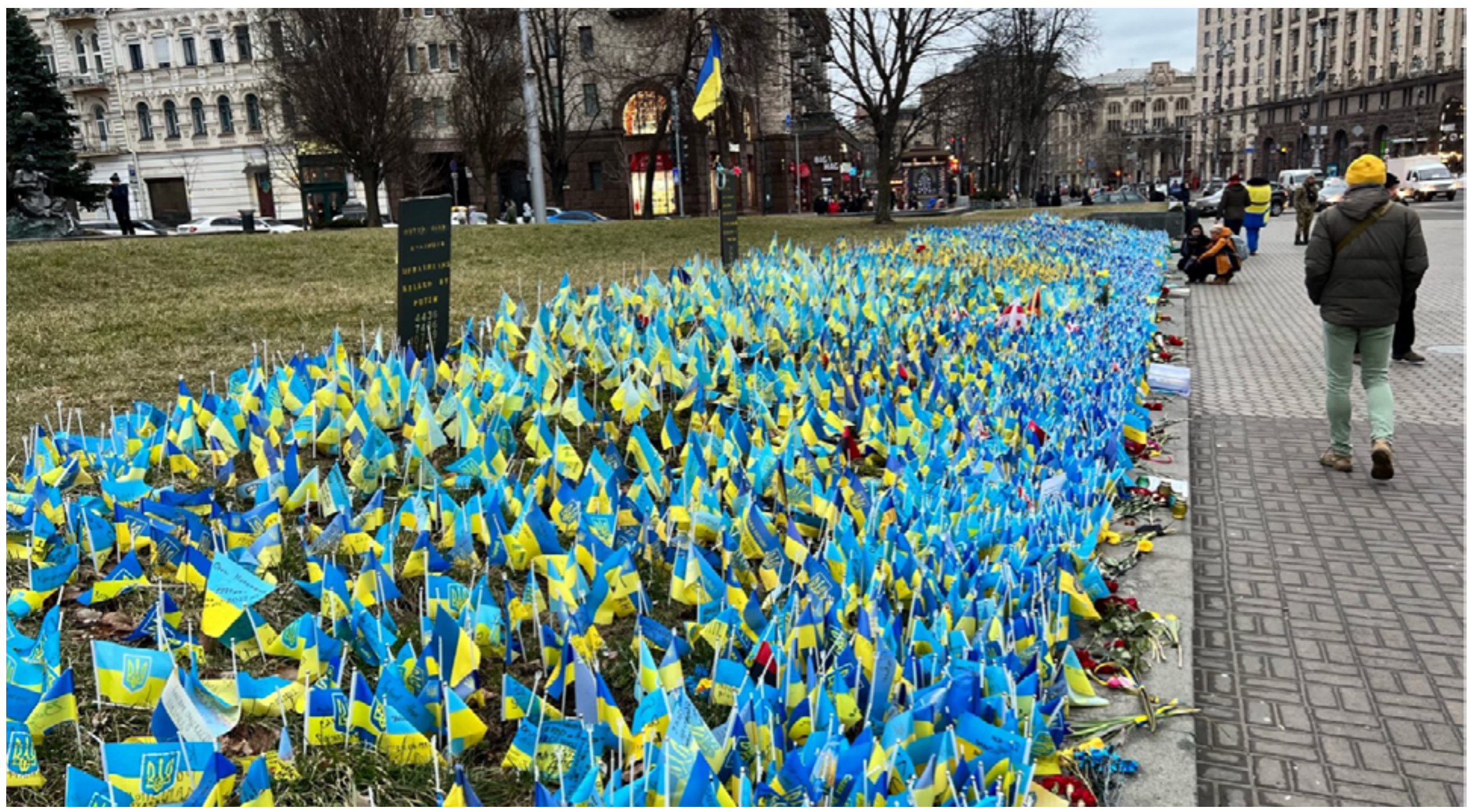

Flags representing fallen soldiers, placed in Maidan Nezalezhnosti, Kyiv, Ukraine. © IOM 2023/Brian MCDONALD

## Data comparison

To compare estimates of internal displacement derived from the three data sources under analysis, we focus on a set of core indicators that are critical for understanding displacement dynamics and informing humanitarian response (IDMC, 2025a; Housset and Bishop, 2025). Developing key summary indicators are important to inform discrete dimensions of the disaster management and response efforts, such as key impacted geographical areas and the severity of impacts, and how these efforts should adapt in response to changing needs during the evolution of a crisis.

The indicators that we focus include displacement rates, which capture population movement trends (departures and arrivals) and help identify the extent of displacement and the key impacted areas (both origin and destination communities); and return rates, which provide insights into resettlement patterns, the pace of return and the challenges faced by returnees. These indicators are important to inform the appropriate planning and delivery of humanitarian assistance to the places and at the times where this is needed. Displacement distances – which help to illuminate the extent of displacement and the social dynamics underlying movement, such as relocation to familiar areas due to family or social ties – would have been a valuable third statistical indicator for comparison. However, this was not feasible due to the absence of flow estimates from DTM Ukraine for 2022 and the unavailability of GPS MD data beyond the end of that year. By focusing on shared indicators, we enable more cross-validation across datasets, while acknowledging their respective limitations, particularly some of the differing underlying concepts outlined in Table 4. This approach strengthens the evidence base for decision-making in complex and rapidly evolving crisis contexts (see e.g. IOM, 2024a; UNHCR, 2025b).

## Table 4. Definitions and use of displacement-related terms across data sources

| Term | Definition | Use in data in this section |
|---|---|---|
| Internally displaced persons (IDPs) | Persons or groups of persons who have been forced or obliged to flee or to leave their homes or places of habitual residence, in particular as a result of or in order to avoid the effects of armed conflict, situations of generalized violence, violations of human rights or natural or human-made disasters, and who have not crossed an internationally recognized State border. | The IOM RDD explicitly estimates the number of IDPs in Ukraine.<br>Digital nontraditional datasets are also used to derive the number of IDPs. |
| Return | In a general sense, the act or process of going back or being taken back to the point of departure. This could be within the territorial boundaries of a country, as in the case of returning internally displaced persons (IDPs) and demobilized combatants; or between a country of destination or transit and a country of origin, as in the case of migrant workers, refugees or asylum seekers. | The IOM RDD estimates the number of returnees, as does the GPS MD data using mobile phone movements, though exact definitions differ significantly as to what is counted as a return.<br>The results in this report show returnee stocks from the IOM RDD and returnee flows from the GPS MD. Estimates of returnees derived from Meta MAPI are not currently available. |
| Migrant stock | For statistical purposes, the total number of international migrants present in a given country at a particular point in time who have ever changed their country of usual residence. | In this section, the migrant stock concept is applied to IDPs rather than international migrants. The IOM RDD collects IDP stocks explicitly; the GPS MD data estimates these from mobile phone locations and past movements; the Meta MAPI data derives IDP stock data from changes in net population stocks over time. |
| Migrant flow | The number of international migrants arriving in a country (immigrants) or the number of international migrants departing from a country (emigrants) over the course of a specific period. | The IOM RDD and GPS MD provide flow data.<br>The former does this from survey responses on home and current location. The latter does this through the movement of mobile devices between areas. |

*Note*: The above definitions are sourced from the *Glossary on migration* (IOM, 2019).

**Interpretation of figures 6, 7, 8 and 10**

The numbers in millions presented for IOM RDD data were extrapolated from percentages to enable comparability across datasets. These figures should now be considered historical, as Ukraine's baseline population has since been recalculated. Beginning with Round 14, population estimates are based on the United Nations Population Fund (UNFPA) baseline for Ukraine, valid as of July 2023 (estimated total population: 33 million, excluding the Autonomous Republic of Crimea and the city of Sevastopol, Ukraine, temporarily occupied by the Russian Federation) (IOM, 2024a).

## 3.3 DISPLACEMENT TRENDS

### IDP national comparison

We begin by comparing national displacement estimates from our three data sources: IOM RDD, GPS MD and META MAPI. Unlike IOM RDD and Meta MAPI, the GPS MD source includes estimates for the Autonomous Republic of Crimea and the city of Sevastopol, Ukraine, temporarily occupied by the Russian Federation. To address this difference, we present national-level figures both including and excluding the Autonomous Republic of Crimea and the city of Sevastopol, Ukraine, temporarily occupied by the Russian Federation (Table 5). For regional-level comparisons, we focus on estimates excluding the Autonomous Republic of Crimea and the city of Sevastopol, Ukraine, temporarily occupied by the Russian Federation to ensure greater comparability with the IOM survey and the estimates produced by Leasure et al. (2023), for which data on the Autonomous Republic of Crimea and the city of Sevastopol, Ukraine, temporarily occupied by the Russian Federation and other Russian Federation-occupied areas prior to the 2022 invasion are not available.

As illustrated in Figure 6, national-level estimates show a relatively high degree of correspondence across data sources, indicating consistency between datasets. Across national and oblast levels, the correlation between GPS MD estimates and Meta MAPI and IOM RDD ranged from 0.36 to 0.96 for Pearson's correlation and from 0.43 to 0.97 for Spearman's correlation (Iradukundu et al., 2025). However, examining percentage differences between the estimates reveals some notable discrepancies in Table 5. Except for a peak observed in July, GPS-based estimates remained consistently lower than the IOM baseline by approximately 5 to 38 per cent across most months.[4] This suggests an underestimation in the GPS MD when compared to the IOM survey. Estimates derived from Meta MAPI are substantially lower than those from the IOM survey throughout the entire period, with differences ranging from approximately 10 to 31 per cent. An exception occurs in August, when the GPS estimates surpass the IOM figures by 11.7 per cent, potentially indicating either a significant shift in the underlying population or a variation in data capture during that month.

[3] Internal DTM Ukraine flow data in percentage format were produced for 2022 but were not made publicly available. Comparisons with GPS MD are also challenging due to the small percentage values involved and associated margins of error.

[4] The largest differences between IOM RDD and GPS MD estimates occur in March and April, reflecting the high population movements immediately following Russia Federation's full-scale invasion; differences are expected given the distinct base populations used. From September to December, refugee movement data were available only at a monthly, rather than daily, resolution, making IDP estimates derived from GPS MD less precise.

## Table 5. IDP national comparison

| Month | IOM RDD | GPS MD (All) | % difference | GPS MD (no Crimea) | % difference | Meta MAPI | % difference |
|---|---|---|---|---|---|---|---|
| February | NA | 3 436 536 | NA | 2 630 000 | NA | 1 003 793 | NA |
| March | 6 478 000 | 6 191 498 | −4.4 | 4 862 000 | −24.9 | 4 931 173 | −23.9 |
| April | 7 707 000 | 7 389 048 | −4.1 | 5 913 000 | −23.3 | 5 304 823 | −31.2 |
| May | 7 134 000 | 7 526 610 | 5.5 | 6 062 000 | −15.0 | 5 226 318 | −26.7 |
| June | 6 275 000 | 7 234 083 | 15.3 | 5 940 000 | −5.3 | 5 669 006 | −9.7 |
| July | 6 645 000 | 6 930 409 | 4.3 | 5 451 000 | −18.0 | 5 766 765 | −13.2 |
| August | 6 975 000 | 9 273 980 | 33.0 | 7 789 000 | 11.7 | 5 759 233 | −17.4 |
| September | 6 243 000 | 5 876 446 | −5.9 | 5 302 000 | −15.1 | 5 692 868 | −8.8 |
| October | 6 540 000 | 4 183 380 | −36.0 | 4 087 000 | −37.5 | 6 099 604 | −6.7 |
| November | 5 914 000 | 3 857 171 | −34.8 | 3 631 000 | −38.6 | 6 538 095 | 10.6 |
| December | NA | 6 748 963 | NA | 6 258 000 | NA | 7 082 230 | NA |

## Figure 6. IDP national comparison

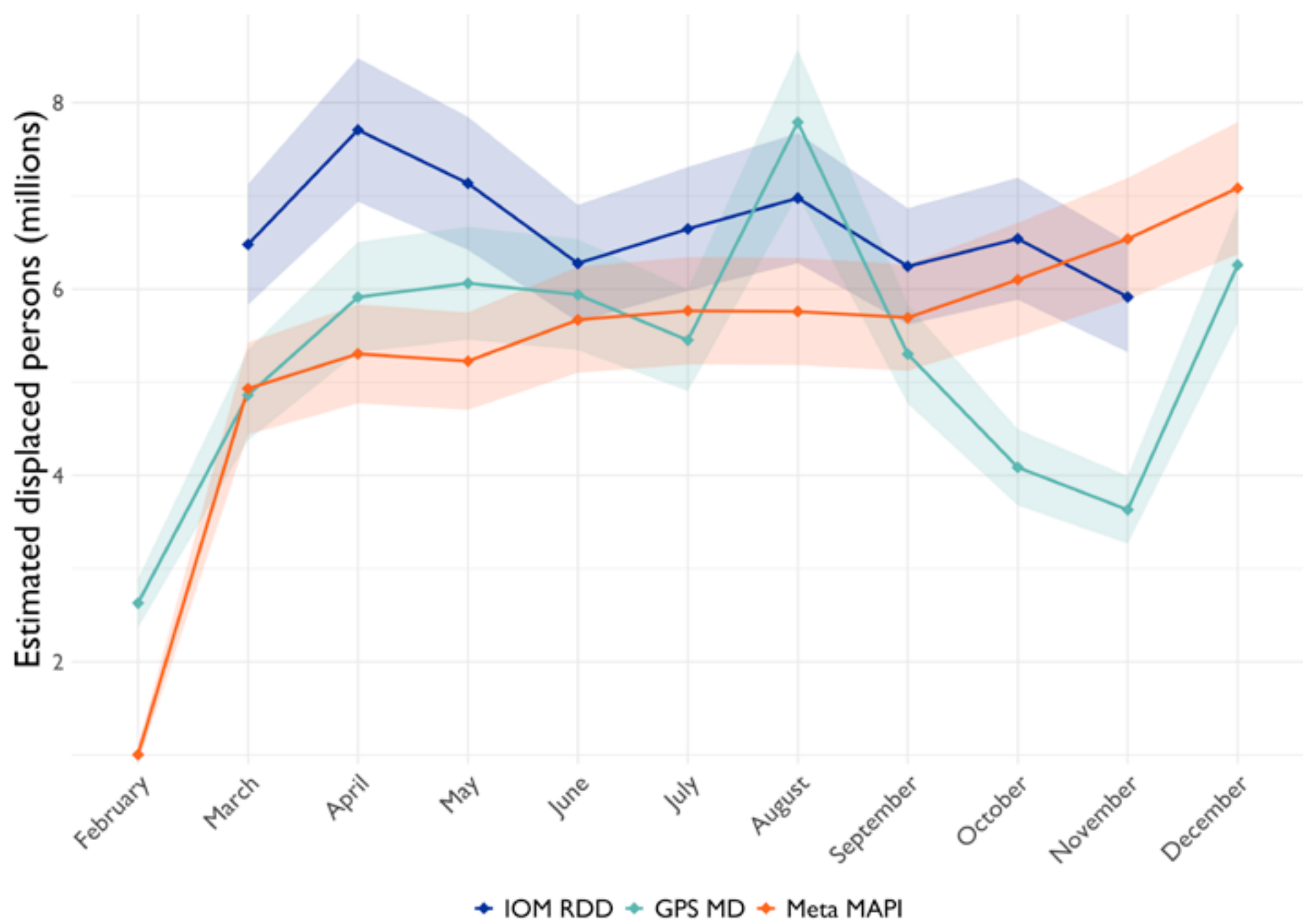

*Notes:* a. IOM RDD: refers to the IOM General Population Survey in Ukraine, a repeated representative sample survey conducted using Random Digit Dialling (RDD) and Computer-Assisted Telephone Interviews (CATI) (IOM, 2024a) (Table 1). The National estimates of rounds 1–11 of the survey are presented here. The population estimates in Figure 6 are no longer used, as those based on the February 2022–May 2023 baseline have been discontinued. Current IDP and returnee figures are derived from survey-based proportions, extrapolated to the UNFPA Common Operational Dataset on Population Statistics (COD-PS), updated in July 2023 with a total population of 33 million (excluding the Autonomous Republic of Crimea and the city of Sevastopol, Ukraine, temporarily occupied by the Russian Federation) (IOM, 2025a).

b. GPS MD refers to GPS phone data, anonymized geolocation information collected from mobile devices, capturing users' movements over time also described in Table 1. Estimates in Figure 6 do not include the Autonomous Republic of Crimea and the city of Sevastopol, Ukraine, temporarily occupied by the Russian Federation. GPS MD estimates account for daily cross-border movement records. A cumulative net count is calculated by subtracting the number of people entering Ukraine from those leaving the country. For September to December, refugee movement data were available only at a monthly, rather than daily, resolution.

c. Meta MAPI, refers from estimated derived from Meta's advertising platform API provides aggregated, anonymized audience estimates based on user demographics and location, offering a valuable proxy for tracking population distribution and mobility in near real time (Leasure et al., 2023).

**Figure 7. IDP oblast comparison**

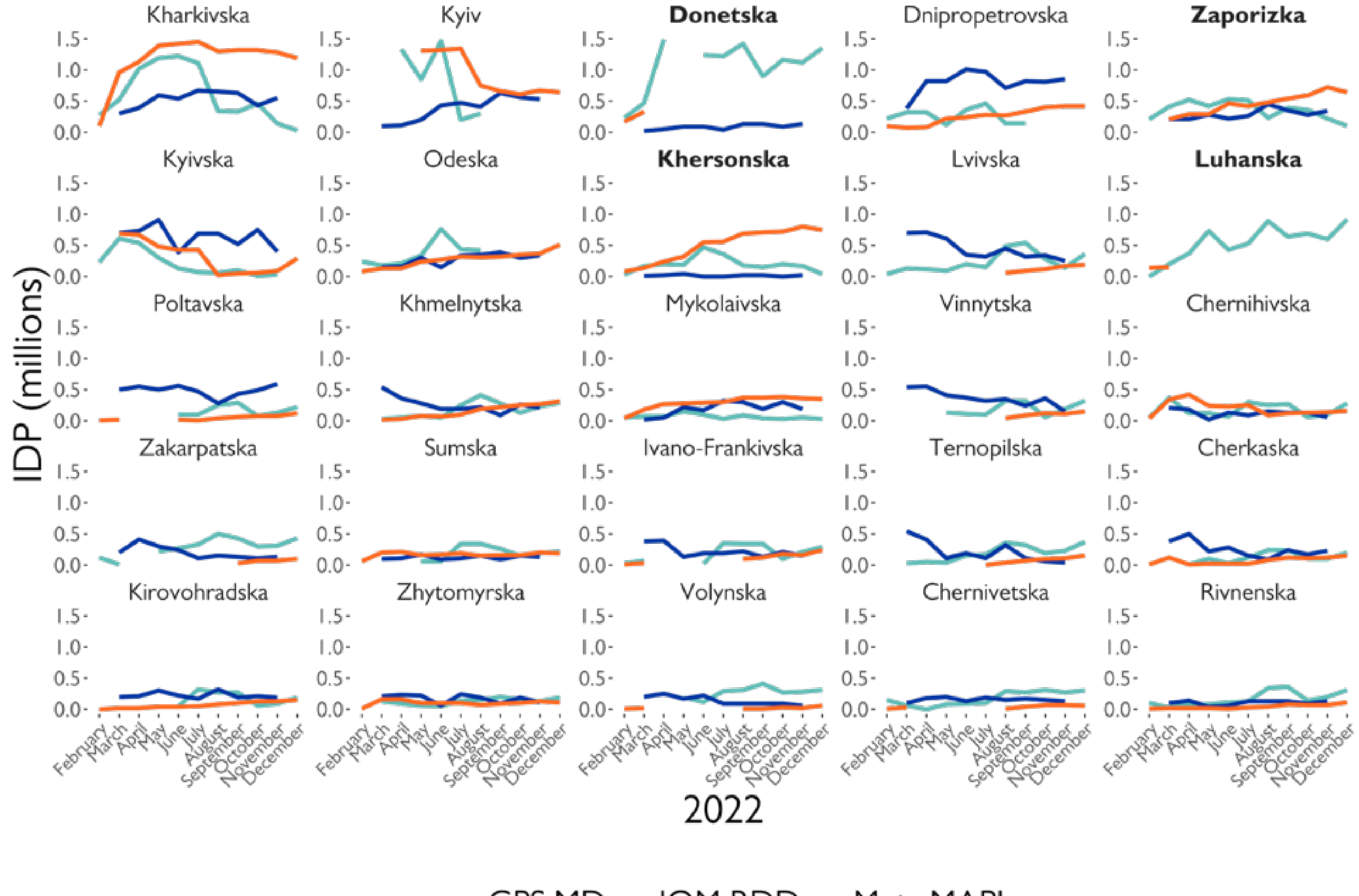

*Notes*: IDP: Internally displaced people in millions. IDPs represents the estimated number of internally displaced persons living in each oblast. Autonomous Republic of Crimea and the city of Sevastopol, Ukraine, temporarily occupied by the Russian Federation is excluded as its data are only available from GPS MD. GPS MD: Estimates derived from GPS Phone Data. Meta MAPI: Estimates derived from Meta's Platform Marketing API. IOM RDD: Estimates from IOM DTM's General Population Survey Ukraine. Data from the IOM RDD survey are based on respondents' self-reported current location, which may impact spatial precision (IOM, 2025a). The numbers in millions presented for IOM RDD data were extrapolated from percentages to enable comparability across datasets. IOM RDD estimates in Donetska, Zaporizka, Luhanska and Khersonska oblasts are underestimated as a result of survey coverage being limited to government-controlled areas and limited numbers of respondents being reached in occupied areas. Oblasts in bold partially occupied by the Russian Federation.

Overall, while GPS MD estimates generally underestimate the displaced population compared to IOM RDD, their deviations are smaller the first six months of the full-scale invasion but become more varied in the second half of 2022. In contrast, Meta MAPI estimates remain relatively stable throughout, suggesting that fluctuations in the number of detected devices affect the GPS data more, whereas social media user counts show greater consistency despite ongoing conflict. However, this apparent overall alignment conceals significant spatial and temporal variability. The geographic distribution of discrepancies indicates that certain oblasts or regions experience more pronounced divergences, likely due to differences in data coverage, survey participation, or local mobility patterns that are not uniformly captured across data sources. While the general alignment supports confidence in comparability, the spatial patterns of these differences require closer examination. Understanding where the largest divergences arise is crucial for accurately interpreting population movements and ensuring such data triangulation exercises can appropriately support humanitarian response efforts.

### IDP regional comparison

Figure 7 and 8 presents oblast-level estimates of internal population displacement based on IOM RDD, GPS MD and Meta MAPI. The oblast is the most granular geographic level at which displacement data is consistently available over the full period of analysis. Although the DTM began producing raion-level (admin 2) estimates in July 2024, our analysis focuses on the short-term period following the escalation of the war in February 2022. We present oblast-level (admin 1) estimates as the primary basis for comparison to maintain consistency across sources.

We assess displacement estimates along two primary dimensions: (1) overall magnitude of displacement, expressed in the number of displaced persons and (2) temporal patterns over the observation period. Across the three data sources – GPS-based mobility data, Facebook-derived estimates and IOM-DTM – generally exhibit alignment in both scale and temporal trends, though some discrepancies emerge upon closer examination. In many instances, GPS-based and Facebook-based estimates show a high degree of correspondence (Figure 7 and 8). The highest displacement levels are observed in Kharkivska, Kyivska, Donetska (based solely on GPS data) and Zaporizka. The latter two oblasts were located near advancing Russian Federation forces during the early stages of Russia's full-scale invasion of Ukraine. Significant displacement in Luhanska is identified exclusively through GPS-based data, likely reflecting gaps in coverage within the Facebook and IOM-DTM datasets. Large movements were also detected in Kyivska and Kyiv City during the Russian Federation's full-scale invasion between February and April 2022. GPS data further identifies high displacement in Sumska, another oblast under immediate threat during the initial months of the invasion.

Crucially, the availability of displacement data across the three sources is neither temporally nor geographically consistent. Several oblasts – namely Donetska, Luhansk and Khersonska – underestimate due to the IOM RDD coverage being limited to government-controlled areas, as well as the limited number of respondents reached in these areas. Similarly, Facebook-derived estimates are unavailable for oblasts such as Vinnytska, Volynska and Zakarpatska. This absence is primarily attributed to Facebook user populations falling below minimum inclusion thresholds, compounded in some cases by power outages and connectivity disruptions that hindered data reporting. For detailed information on the data inclusion criteria for Facebook-based estimates, we refer readers to the Supporting information of Leasure et al. (2023) paper on Nowcasting Daily Population Displacement in Ukraine.

These findings reinforce several key observations introduced in the opening section of this report. A major advantage of GPS-based mobile data is its extensive geographic coverage, encompassing areas of the Donetsk, Luhansk, Kherson and Zaporizhzhia regions of Ukraine occupied by the Russian Federation. While the estimates presented here are aggregated at the oblast level, the underlying data can support analysis at a much finer spatial resolution, enabling detailed tracking of displacement patterns (Figure 9). For example, Figure 9 presents raion-level estimates for selected key months within the first six months of the war – granularity not available from Meta or IOM-DTM sources. We focus on March, May and June 2022, as each corresponds to notable displacement events (Walker, 2025). March captures the period around 14 March, when initial evacuations occurred, particularly from Kyiv City, Kharkivska and Kyivska oblasts.

**Figure 8. Temporal mapping of IDP patterns in select months of 2022 at oblast level**

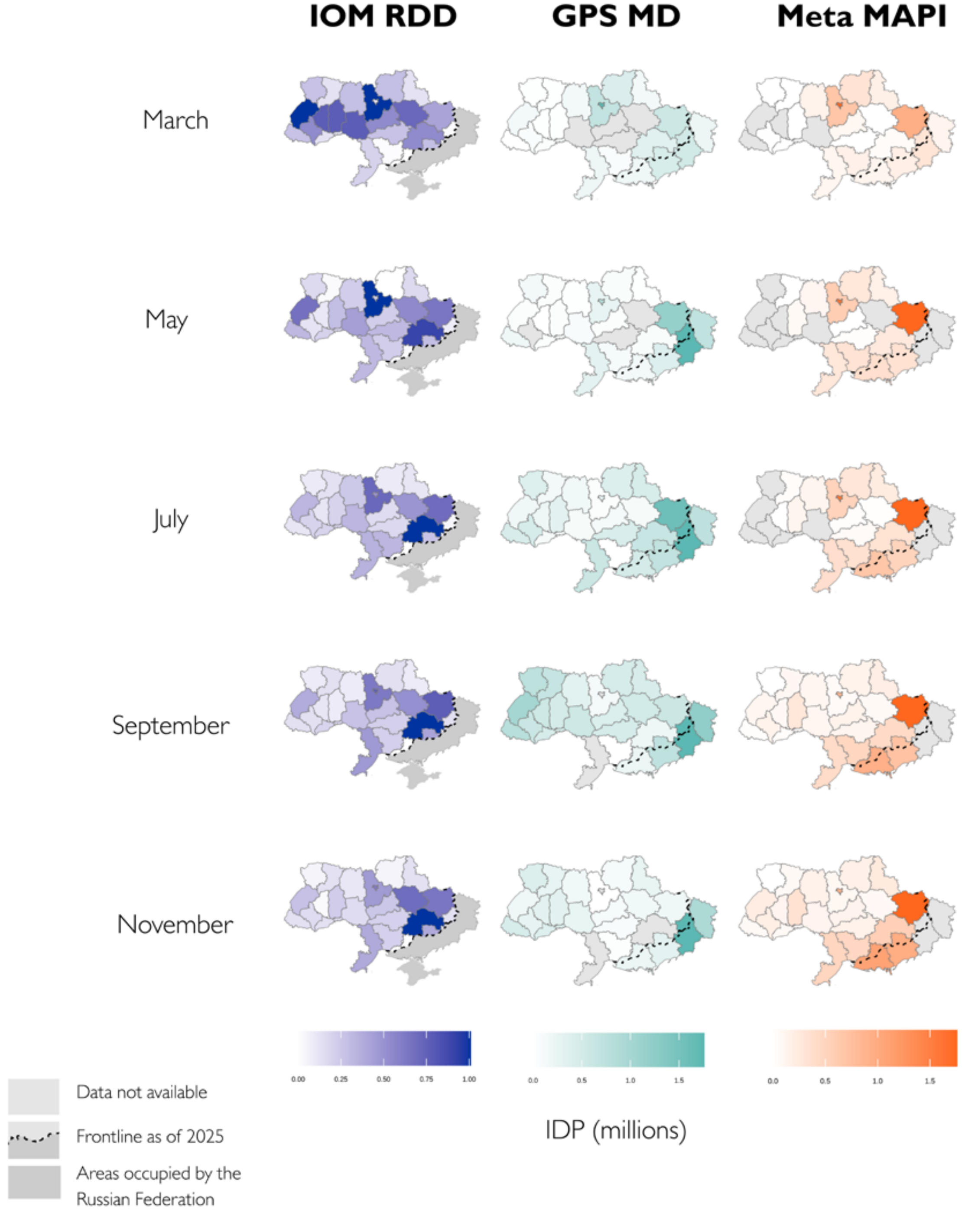

*Source:* Elaborated by the authors using R.

*Notes:* IDP: Internally displaced people in millions. IDPs represent the estimated number of internally displaced persons living in each oblast. The Autonomous Republic of Crimea and the city of Sevastopol, Ukraine, temporarily occupied by the Russian Federation. is excluded as its data are only available from GPS MD. GPS MD: Estimates derived from GPS Phone Data. Meta MAPI: Estimates derived from Meta's Platform Marketing API. IOM RDD: Estimates from IOM DTM's General Population Survey Ukraine. Data from the IOM RDD survey are based on respondents' self-reported current location, which may impact spatial precision (IOM, 2025a). The numbers in millions presented for IOM RDD data were extrapolated from percentages to enable comparability across datasets. IOM RDD estimates in Donetska, Zaporizka, Luhanska and Khersonska oblasts are underestimated as a result of survey coverage being limited to government-controlled areas and limited numbers of respondents being reached in occupied areas. This map is for illustration purposes only. The boundaries and names shown and the designations used on this map do not imply official endorsement or acceptance by the International Organization for Migration.

**Figure 9. Snapshot mapping of GPS MD at raion-level 2022**

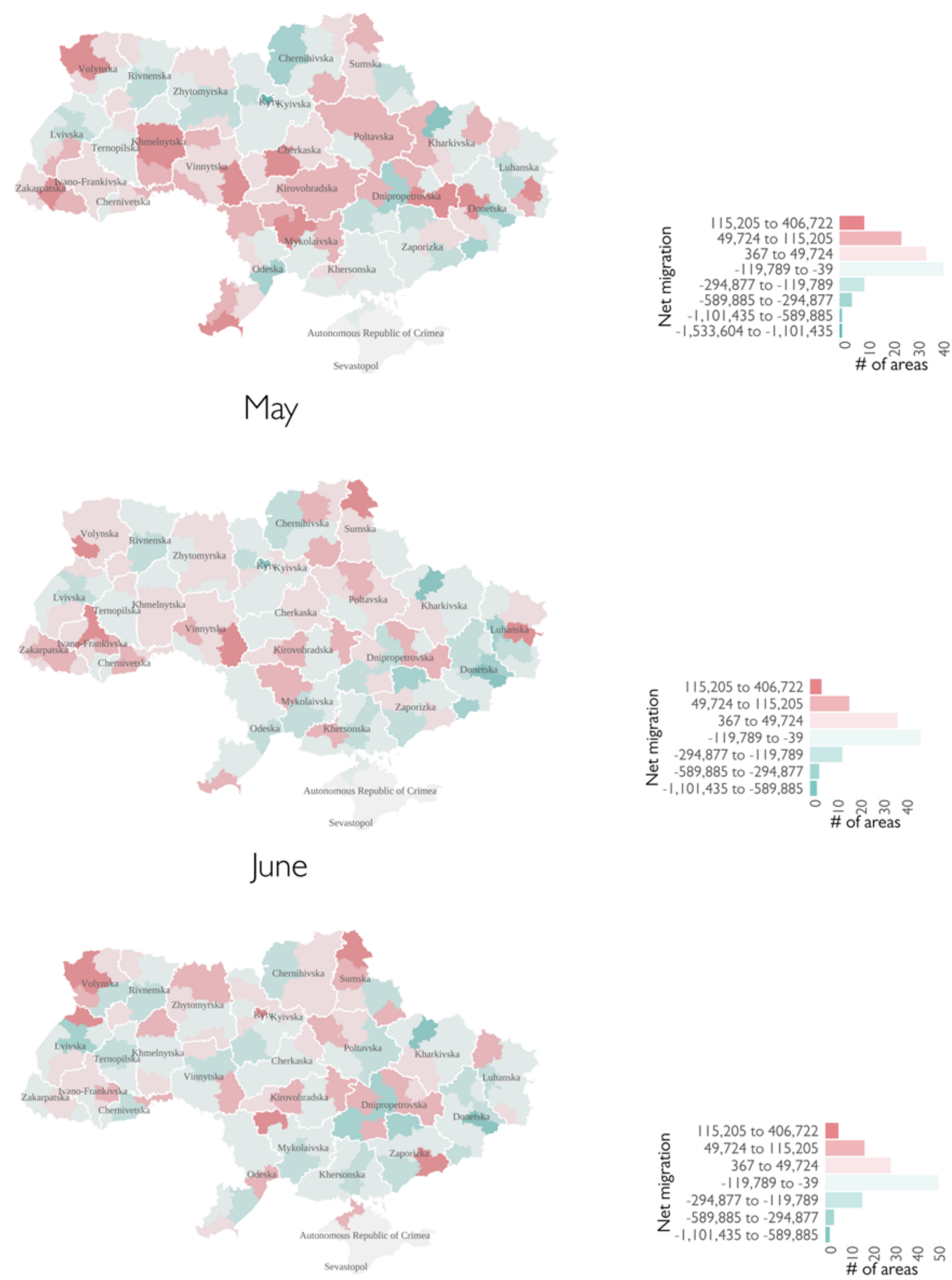

*Source:* Elaborated by the authors using R.

*Note:* These maps are for illustration purposes only. The boundaries and names shown and the designations used on this map do not imply official endorsement or acceptance by the International Organization for Migration.

May reflects movements surrounding the first evacuation of Khersonska oblast on 14 May, as well as the return of nearly half a million people during Orthodox Easter on 24 April. June captures the period of the second evacuation of Khersonska oblast on 15 June. While the maps are presented at the raion level, even finer spatial resolution is possible but not included here.

The comprehensive reach and granularity are particularly valuable during the initial weeks following the onset of the conflict or a sudden-onset disaster, when timely and granular information is critical. Facebook-based displacement estimates offer comparable advantages, although their spatial resolution is generally coarser and data coverage is incomplete across all oblasts. However, Facebook data provides additional value by offering limited demographic insights, which can help to better understand the composition of displaced populations and the characteristics of those on the move.

Once IOM-DTM estimates become available, they tend to align closely with trends observed in the GPS and Facebook data. Notable discrepancies, however, appear in oblasts such as Donetska, given IOM RDD is limited to government-controlled areas. Similar inconsistencies are observed in Kharkivska and Khersonska. In the case of Kyiv City, displacement estimates derived from both GPS and Facebook data are significantly higher than those reported by IOM. This may reflect methodological differences in how displacement is captured in dense urban centres and during periods of rapid movement, and is suggestive of possible underreporting in the IOM RDD data.

In looking at finer-grained, raion-level data for Ukraine – which can, in principle, be derived from GPS mobile phone data aggregated at much higher spatial resolution – it is important to note the potential need for collaboration with government entities. GPS tracking of individual mobile devices is not officially conducted, as geolocation data are considered personal under the Law of Ukraine "On Personal Data Protection" (2010) and the Law of Ukraine "On Electronic Communications" (2020) (RADA, 2010). Even the State Statistical Service is not authorized to access such information in recent years. As a result, any use of digital trace data at this level must carefully consider legal restrictions and the role of relevant authorities to ensure compliance with national regulations.

### 3.4 COMPARING RETURNEES

Figure 10 presents monthly oblast-level estimates of returnees based on GPS MD and IOM RDD. Data on returnees can be derived throughout the entire period of analysis from the GPS-based mobility data. In contrast, DTM Ukraine began reporting the share of returnees by oblast after April 2022, with the estimated number of returnees per oblast provided only after the 14th round of the IOM RDD that reported data for September 2023. The numbers in millions presented for IOM RDD data were extrapolated from percentages as reported by IOM-DTM to enable comparability across datasets.

Due to differences in data-collection methodologies, the estimates plotted in Figure 10 are not directly comparable. Whereas the GPS MD shows the flow of returnees into an oblast, the IOM RDD displays the stock of returnees by oblast for each time interval. Furthermore, the way in which each dataset defines returnees differs. Whereas the IOM defines a returnee as

**Figure 10. Returnee oblast comparison February-August 2022**

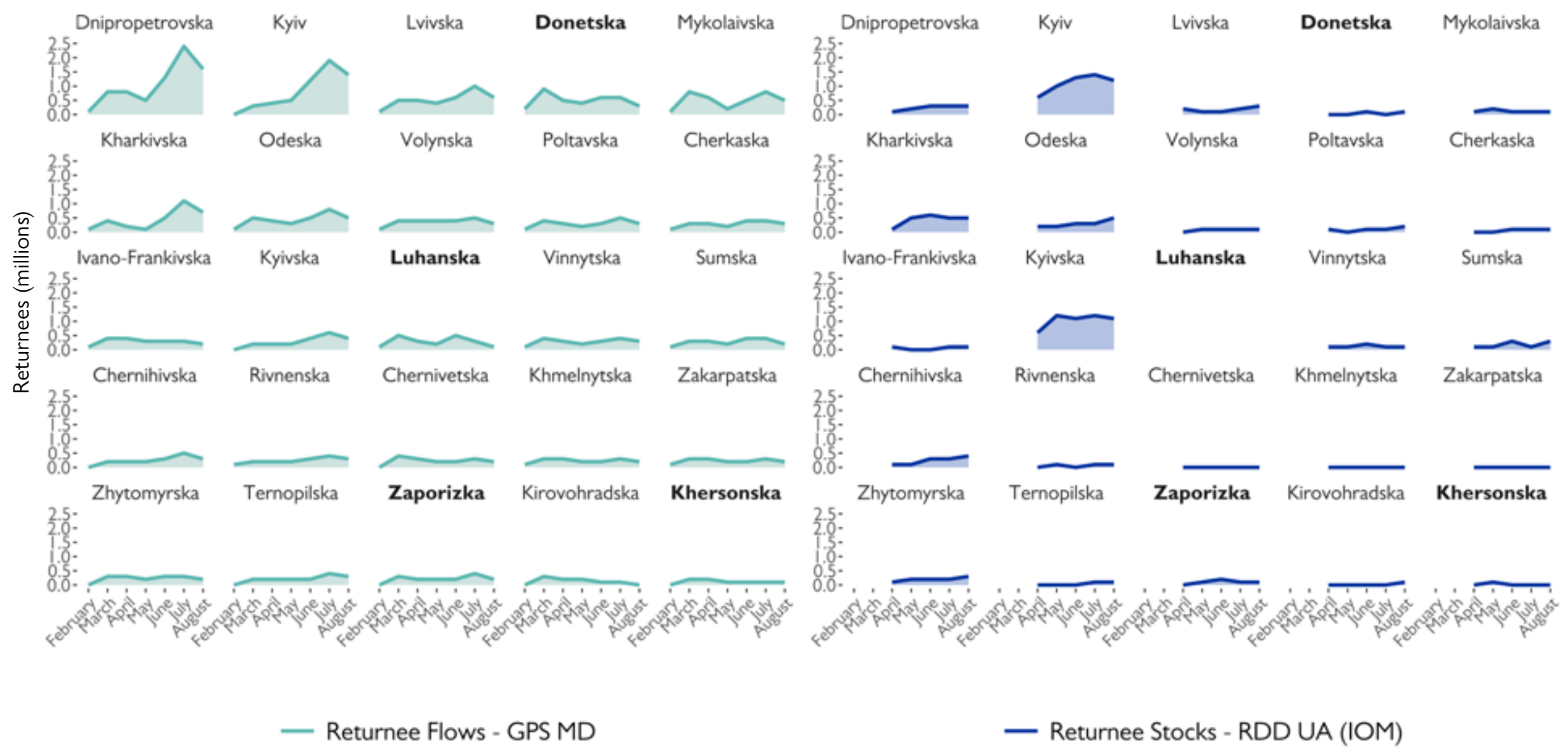

*Note*: IOM RDD collects data based on respondents' current location and relies on self-reported information, which may affect spatial precision (IOM, 2025a). The Autonomous Republic of Crimea and the city of Sevastopol, Ukraine, temporarily occupied by the Russian Federation is excluded from this analysis because data for this region are available only from the GPS MD source. Oblasts in bold text are those either fully or partially occupied by the Russian Federation during the data-collection period. As a result, selected raions within these oblasts were excluded from the IOM RDD data collection.

someone who left their habitual residence for at least two weeks and has since returned, applying this definition to the GPS MD yielded unrealistic results, as outlined in Iradukunda et al. (2025). Instead, using a similar definition to Leasure et al. (2023), the GPS data defined a return as a device which was recorded away for an average of nine weeks and subsequently in the same area that was identified as home location before the start of the full-scale invasion.

These definitional differences highlight some difficulties in synthesizing traditional data streams with digital ones. In this case, much of the comparability issue is due to the definition of a return not aligning, rather than the data itself; however, these definitions are shaped by what each data source captures. Indeed, ensuring definitions are consistent between data is always a challenge, but between survey and digital nontraditional data these difficulties can be magnified by the differences in exactly what is being measured. The GPS mobility data, for example, captures the location of mobile devices rather than an individual themselves. This means that, because devices may return one or more times to a location, estimating the total number of returnees at one point in time can present challenges. In contrast, the IOM RDD data can effectively capture stock data from survey rounds but obtaining data on the flow of returnees from one area of Ukraine to another at a detailed spatial level is difficult given limited sample sizes. These difficulties also demonstrate how different data streams can complement each other and where digital trace data can add value to traditional sources. In terms of temporal and spatial resolution, for example, the GPS MD can provide a precise time point for when the return occurred at a granular spatial level, as well as information on returnee movements within Ukraine at a similarly detailed temporal and spatial level. Further, whilst multiple returns with the same device causes some issues in estimating stocks, it is an additional and potentially helpful data point to understand cases where individuals experience multiple displacement and return events over the course of a crisis period. Reliance on mobile devices means that when these devices are replaced or fall out of use, the associated data points are lost from the dataset, making it more difficult to effectively track and capture information on returnees.

As highlighted in the recommendations section of this report (Figure 5), it is essential that the integration of digital sources into existing humanitarian data systems fosters trust, employs transparent methodologies and supports effective use. The challenges related to returnee data, as outlined above, underscore the need to carefully synthesize diverse data streams, acknowledging the unique strengths and limitations of each. Looking ahead, greater efforts are required to enhance the comparability of returnee datasets, thereby providing policymakers and humanitarian actors with the confidence that digital nontraditional data can be both reliable and actionable in this context.

# CHAPTER 4

## INSIGHTS FROM PAKISTAN: EXPANDING THE USE OF TRIANGULATED DATA FOR HUMANITARIAN RESPONSE

Building on the work triangulating data to quantify conflict-induced population displacement in Ukraine, the project's third workshop examined how similar methodologies could be scaled and adapted to a wider range of disasters and operational contexts. Ideally, the rapid triangulation of diverse data sources should be applicable across crises driven by conflict, climate-induced natural hazards and epidemics. To test this potential, the third workshop took the form of a hackathon – designed to further explore how traditional and digital nontraditional data can be integrated to support more timely and effective humanitarian response in disaster settings.

The hackathon was supported by Snowflake's AI Data Cloud – a silo-free environment where you can collaborate over data with your teams, partners and customers and even integrate external data, models and applications for fresh insights – specializing in data warehousing and analytics. Snowflake provided the infrastructure to store and manage the datasets used during the event and contributed demonstrations and code samples to facilitate data processing, visualization and analysis – enabling participants to work efficiently with complex and diverse data sources.

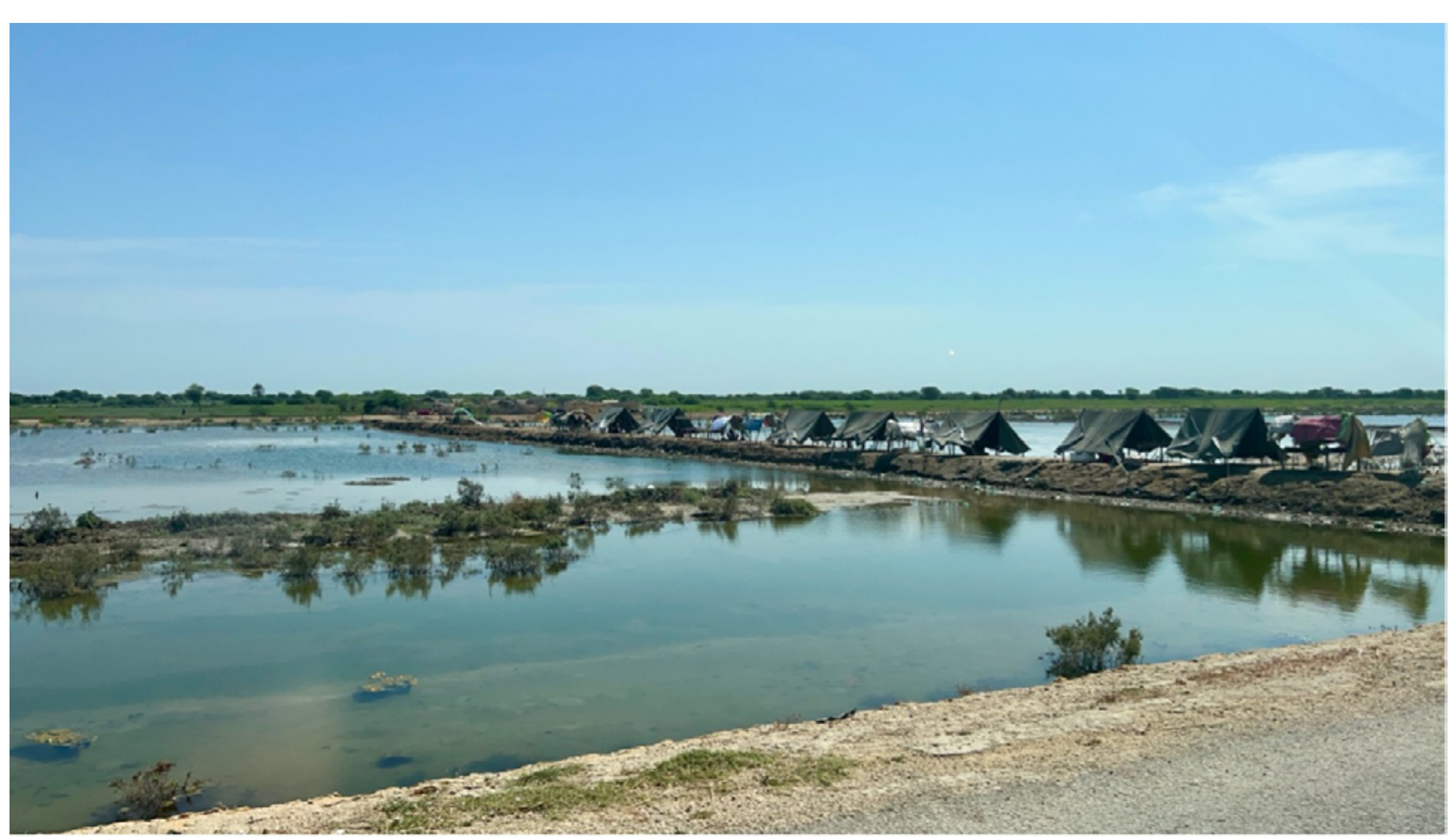

IDPs displaced close to the Hyderabad-Badin provincial highway, Sindh, Pakistan. © IOM 2022/Brian MCDONALD

## 4.1 PAKISTAN FLOODS 2022: CONTEXT AND DATASETS

The hackathon focused on the 2022 floods in Pakistan as a case study. Between June and October 2022, unprecedented monsoon rains affected over 33 million people, resulting in 1,739 deaths and an estimated USD 40 billion in damages (Rowe, 2022). Participants were provided with a diverse set of datasets, including data from the DTM's Community Needs Identification (CNI PK) survey – a village-level key informant tool used to estimate displacement and assess the multisectoral needs of flood-affected communities (Table 2) (IOM, 2023). They also worked with digital nontraditional datasets from Meta's Data for Good (Meta DfG) initiative, namely "Facebook Population During Crisis" and "Facebook Movement During Crisis," offering stock and flow data on user presence and mobility patterns at low temporal and spatial resolution (Table 2). These were complemented by data on flood severity and extent, contextual demographic information, relative deprivation indices and climate indicators – together providing a comprehensive, multi-layered view of the crisis to support data-informed humanitarian decision-making.

Full details of the hackathon, including data access, methodology and participant resources, can be found at: https://pietrostefani.github.io/pop-displacement-disaster/.

## 4.2 KEY CONCLUSIONS FROM THE HACKATHON

The key conclusions from the hackathon evidence and expand key points outlined in Section 1 of the report (see Figure 2). The hackathon highlighted the value of integrating traditional and digital nontraditional data streams, providing practical examples of how this synthesis can be achieved. It underscored the importance of leveraging each data source to offset the limitations of others, demonstrating how combining datasets can generate richer, more actionable insights.

Our findings emphasized the need for clear visual outputs and well-documented analytical pipelines and their development to ensure replicability for future crisis scenarios and rapid response. Overall, the hackathon showcased the potential of digital nontraditional data to enhance traditional humanitarian data systems, offering new opportunities to inform decision-making and shape effective policy.

### Table 6. Key conclusions from the hackathon

| | |
|---|---|
| **Drawbacks of traditional data sources** | The IOM data provided valuable qualitative insights into community needs and offered an indication of the scale of displacement resulting from the disaster. However, the timing of data collection – conducted after the initial major flooding events – was influenced by the need to recruit and train enumerators and to complete baseline assessments used to select survey locations. **Findings of community needs initiatives (CNI) are therefore often not fully generalizable**, as the assessment cover only selected villages. This limitation was particularly pronounced in the earlier rounds of data collection, when coverage was more limited. |
| **Where digital nontraditional data can (and cannot) fill the gaps** | **Digital nontraditional data can provide wide spatial and temporal coverage and high-resolution data**. These data can support broader geographical coverage from the onset of a crisis by providing temporally and spatially granular information across affected regions, as well as pre-crisis baseline data to enable comparisons over time. However, their generalizability may be **limited in contexts with low social media penetration**, which can affect the representativeness of the insights derived. Furthermore, **digital trace data from GPS technology does not provide the detail in terms of displaced populations' demographics or characteristics** that the CNI could. |
| **Combining traditional and digital nontraditional data is not always simple** | **Digital nontraditional data is often available at different geographic units than traditional data**. Geographically joining these datasets is therefore challenging. As a result, **synthesizing traditional and digital nontraditional data often required additional processing** to ensure geographic scales are matched up. |
| **The optimal spatial unit for analysis depends on context** | **Different research questions and use cases required outputs to be at different spatial scales** to ensure the insights gained from them were as useful as possible. Whilst digital trace data is often very fine-grained spatially, to answer questions related to communities, these data had to be **aggregated up to identify areas experiencing higher levels of displacement** and to **prioritize locations where larger movements were occurring** and populations were most in need. Similarly, the settlement/village level data was not suited to answering country-level questions during the crisis period. **These challenges highlight the need for flexible, multi-scale data systems that can adapt to diverse information needs – ranging from localized targeting to national-level planning – while ensuring coherence across spatial levels**. |
| **Clear visualizations, insights and communication are vital** | Clear visual representation and concise messaging are essential for effective communicating complex outputs and enabling practical action. **Prioritizing digestible, easy-to-understand formats** is particularly important to support decision-making by policymakers and humanitarian actors who may not have data-specific expertise. This also facilitates the r**apid production and use of outputs in fast-moving crisis contexts**, where timely and actionable insights are critical. |
| **The importance of clarity around data sources and their limits** | The importance of clarity around data sources and their limits: Traditional and digital nontraditional datasets both come with advantages and challenges that should be acknowledged. **Transparency around the strengths and weaknesses of each dataset adds trust and enables data insights to be more accurately interpreted**. |

## 4.3 PROBLEM STATEMENTS

The hackathon saw participants split into four groups, each reflective of the stakeholders and organizations represented at the event. Each group focussed on one of four problem statements that focused on a specific area of research. These are displayed in the table below.

**Table 7. Problem statements used by groups during the hackathon**

| **Event-based or rapid response: Developing key indicators** | **Learning from the past: Patterns and resilience** |
|---|---|
| • Focus on generating timely, actionable indicators from Facebook and IOM data to support immediate disaster response operations.<br>• Develop key indicators such as displacement rates, volume of movement, identification of vulnerable populations or most affected regions.<br>• Explore how these indicators compare with – or can be enriched by – IOM data collected in reference to the 2022 flooding event in Pakistan.<br>• What aspects of local population, their movements and geographical context are most useful to humanitarian responders in the first hours and days of a disaster? | • Analyse data to identify behavioural trends and resilience patterns from the 2022 floods to present (IOM Pakistan Flood Responses Rounds 1–6).<br>• Are there repeated behavioural patterns in terms of population movement, key origins, key destinations and population distribution across different flood events in Pakistan?<br>• Which regions show higher levels of resilience and what factors contribute to that resilience (e.g. infrastructure, demographics, preparedness)?<br>• Use comparative analysis to highlight learnings that can inform future planning and interventions. |
| **Predicting future movement: Mobility modelling** | **Relationship between flood severity and extent of displacement** |
| • Develop predictive models to forecast displacement patterns in future disaster scenarios by leveraging social, economic and geographic indicators, using data from the International Organization for Migration (IOM) and Facebook (Meta) mobility data.<br>• What kinds of areas are most likely to receive displaced populations? (more vulnerable communities?).<br>• Can we identify a systematic set of predictors for where people move from and to during disasters?<br>• How can the information extracted from these analyses be used to improve preparedness and resource allocation in high-risk regions and destination community areas. | • How are flood severity metrics associated with higher displacement volumes or longer displacement durations?<br>• Can threshold points in flood impact metrics be identified? That is points beyond which displacement becomes significantly more likely.<br>• How do displacement outcomes differ in high-severity versus low-severity flood zones? |

## 4.4 HACKATHON RESULTS: PROBLEM-SOLVING IN PRACTICE

Groups used the hackathon to answer one of the above problem statements. Each produced a presentation with a variety of visualizations showcasing their results and analysis. Below are some examples of what was produced, as well as brief summaries written by each of the teams around their problem statement, their approach to answering it and their outputs.

**Group 1:** Elisabetta Pietrostefani, PhD (UoL), Ana Valera Valera, PhD (LSE), Hong Tran-Jones (IOM), Nando Lewis, PhD (IOM), Franziska Clevers (IOM), Lorenzo Sileci, PhD (LSE), Flora Chu (Snowflake), Linh Hoang Thuy (Snowflake) and Scott White (FCDO).

### Problem statement: Relationship between flood severity and extent of displacement

Group 1 investigated the relationship between flood severity and the extent of displacement during the 2022 floods in Pakistan, aiming to understand how geospatial indicators could inform more timely and equitable humanitarian responses. Using flood permanence data from the Copernicus Emergency Management Service's (CEMS) Global Flood Monitoring (GFM) system as a proxy for flood severity, they assessed the percentage of valid satellite observations in which flooding was detected between 10 August and 23 September 2022 – capturing the peak of the disaster. This was combined with digital mobility data from Meta's "Facebook Population During Crisis" dataset, which tracked net population changes from 14 August to 7 September 2022 relative to a pre-flood baseline.

By overlaying these datasets using Uber's H3 Level 5 spatial grid (approximately 250 $km^2$ per hexagon), the team examined displacement patterns at a finer resolution than typical administrative boundaries, enhancing the potential for more targeted operational decision-making. Their analysis indicated that areas with higher flood permanence generally corresponded to greater outflows of population, suggesting a meaningful relationship between severity and displacement. However, several limitations were identified: Sentinel-1 satellite data may miss flooding in urban areas; Facebook data only represents app users with connectivity – excluding populations in remote or underserved regions – and is subject to suppression in areas with low user density. Calibration to actual population figures was also not applied and cultural or demographic factors influencing migration decisions were not directly captured.

Despite these limitations, the approach demonstrates strong potential for scalability and real-time applicability. Both the GFM and Meta datasets offer global coverage and could be deployed rapidly in future crises. Moreover, the team suggested that integrating additional layers – such as demographic and socio-economic indicators, or IOM's multisectoral needs assessments – could significantly enhance the analysis. For example, regression modelling could reveal how flood exposure intersects with underlying vulnerabilities, supporting more equitable policy recommendations. In turn, such analysis could help predict not only where people are likely to move but also what their needs may be – enabling faster, more precise and more inclusive humanitarian response strategies.

**Figure 11. Bivariate map of displacement and flooding levels**

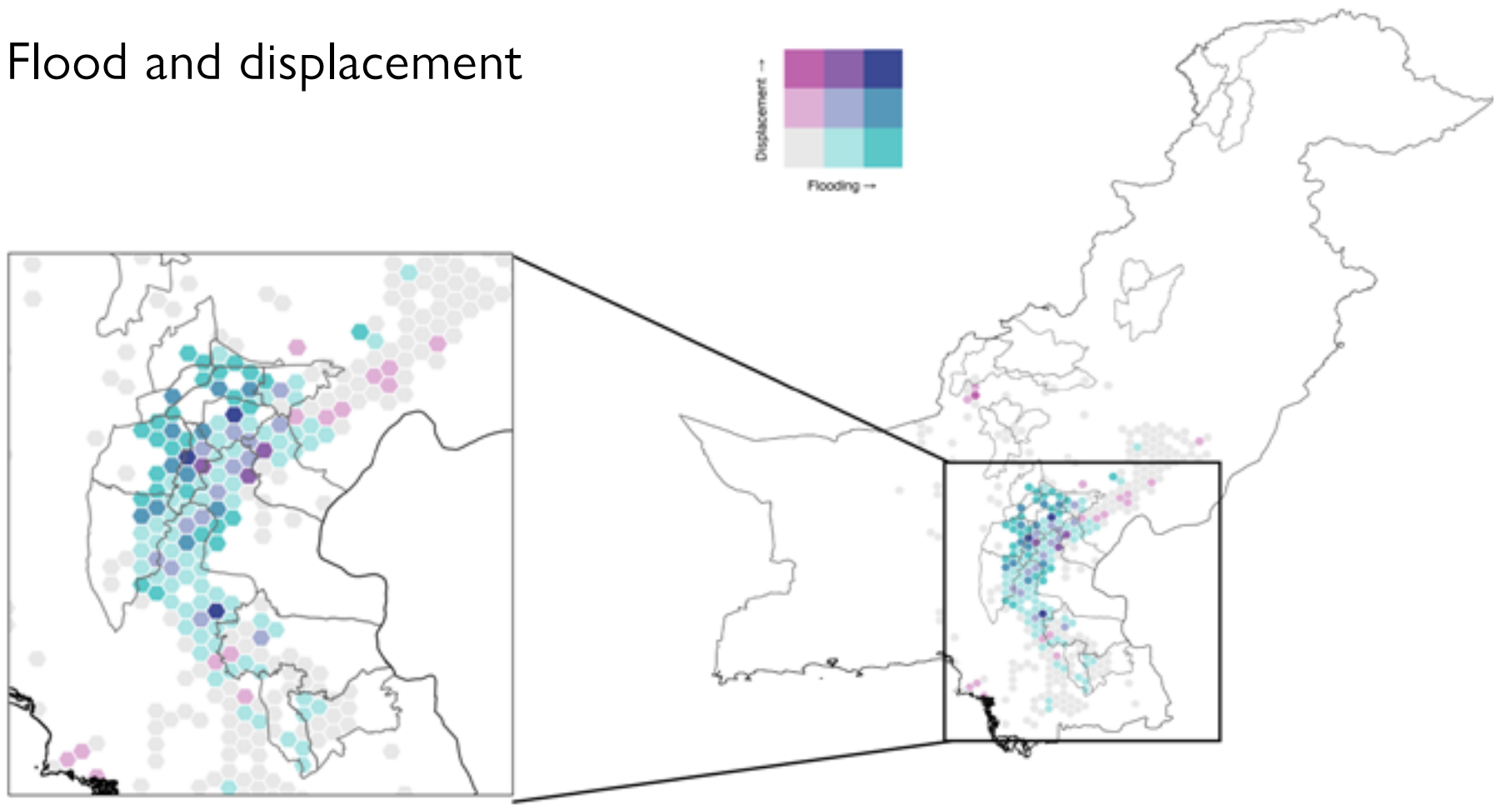

*Source:* Generated by Group 1 using R during the Berlin hackathon and is not a final output but demonstrates work done during hackathon.
*Notes:* This map is for illustration purposes only. The boundaries and names shown and the designations used on this map do not imply official endorsement or acceptance by the International Organization for Migration.
Displacement figures are from the Facebook Population During Crisis data; flooding is computed from Copernicus Emergency Management Service. Polygons are areas where IOM has conducted data-collection exercises.

**Group 2:** Francisco Rowe, PhD (UoL), Andrea Aparicio Castro, PhD (University of Oxford), Dominik Bursy (Humboldt University of Berlin), Joe Slowey (IOM), Laura Coskum (IOM), William Lumala (IOM) and Euan Newlands (Snowflake).

**Problem statement: Learning from the past: Patterns and resilience**

Group 2 focused on the challenge of "Learning from past patterns," aiming to understand which routes and destinations experienced the highest displacement flows during the 2022 floods in Pakistan – and the contextual factors that shaped those movements. The team sought to characterize these patterns by combining traditional and digital trace data sources, identifying the conditions under which certain areas experienced greater inflows or outflows of displaced populations.

This focus was selected for three main reasons. First, it was achievable within the limited timeframe of the 1.5-day hackathon. Rather than building a predictive model or developing entirely new indicators, the team concentrated on comparing two robust, pre-existing datasets: IOM's DTM survey data and Meta's Facebook Movement During Crisis data. This targeted approach enabled the production of a well-defined, actionable output.

Second, the analysis had clear operational relevance. IOM practitioners on the team emphasized that understanding past displacement is often one of the strongest indicators of where people might move in future crises. By anchoring their analysis in Facebook's digital mobility data and contextualizing it with on-the-ground survey data, the team generated insights that could meaningfully inform preparedness and response strategies.

Third, the project illustrated the broader value of integrating traditional survey data with digital trace data – a central goal of the hackathon. By mapping Facebook movement data onto an 800-metre grid, the team enabled a high-resolution analysis of displacement pathways.

The methodology involved several steps. The team linked five rounds of IOM displacement survey data (collected from July to October 2022 at the union council level) to corresponding centroids in Facebook's mobility grid. This allowed them to compare displacement volumes over time. Using the Facebook flows, they extracted and ranked the top ten origin–destination corridors based on net movement. To explore the drivers of these patterns, they added contextual variables – sociodemographic characteristics (e.g. age, gender and deprivation) and environmental factors such as rainfall.

The final output visualized key displacement flows, including colour-coded maps and population profiles for major origin and destination points. The work demonstrated how blending digital and traditional data can generate timely, actionable insights to support more effective disaster preparedness and response.

**Figure 12. Maps of decrease in daily Facebook population and map of deprivation levels in Pakistan**

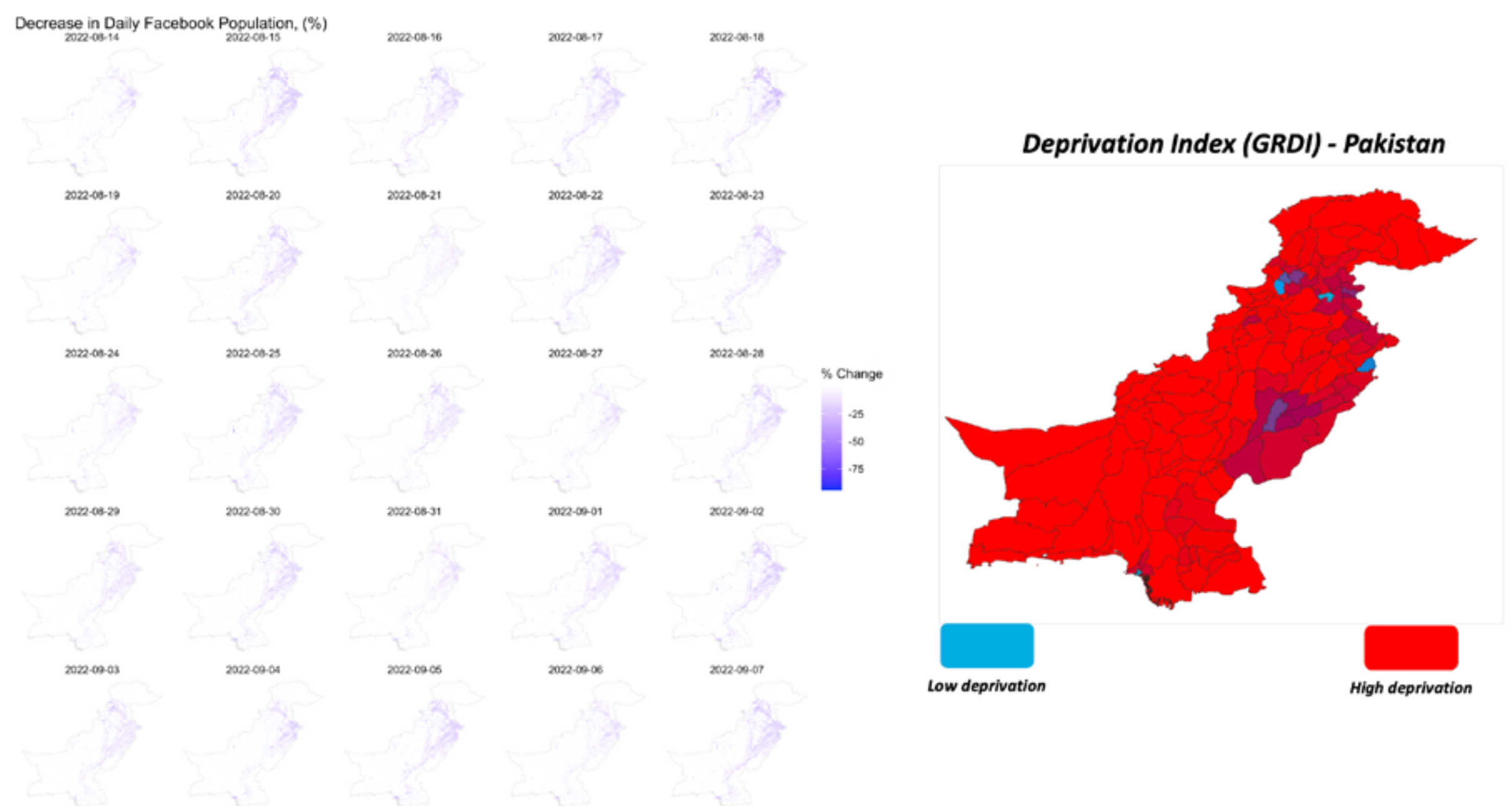

*Source*: Generated by Group 2 using R during the Berlin hackathon and is not a final output but demonstrates work done during hackathon.
*Notes*: The maps are for illustration purposes only. The boundaries and names shown and the designations used on this map do not imply official endorsement or acceptance by the International Organization for Migration.
The decrease in population was taken from the Facebook Population During Crisis data and deprivation data was taken from the Global Relative Deprivation Index.

**Group 3:** Gabriele Filomena, PhD (University of Liverpool), Luong Tran (IOM), Michael Zihanzu (IOM), Huan Wang (IOM), Benjamin Pfau (Snowflake), Kate Hodkinson (UUN OCHA) and Adam Bekele (Snowflake).

**Problem statement: Learning from the past: Patterns and resilience**

Group 3 addressed the challenge "Learning from the past: Patterns and resilience," with the objective of analysing displacement and behavioural trends in response to Pakistan's 2022 floods. Using IOM's Flood Response data alongside Facebook Data for Good, the team aimed to explore how communities responded to and recovered from major flood events. Central to their inquiry were key questions: What patterns emerge in population movements – where do people flee from, where do they go and how do these distributions evolve over time? Which regions exhibit higher resilience, particularly in terms of returns, and what factors – such as infrastructure, demographics or preparedness – support this resilience?

The group placed special emphasis on understanding the dynamics of return: how and when displaced populations come back to their places of origin, and what distinguishes areas with higher return rates or faster recovery. Their methodology sought to leverage digital trace data to capture immediate post-flood movements, with Facebook mobility data offering near-real-time insights into destinations and behavioural responses in the critical days following heavy rainfall. This short-term analysis would then be complemented by longer-term trends using IOM displacement data, to track ongoing movements and eventual returns over several months.

Although time constraints limited the team's ability to fully implement the planned analysis, their intended approach included producing visual representations of displacement flows, identifying persistent corridors of movement and mapping return behaviour. Resilience was conceptualized not only through return rates but also through contextual characteristics – such as proximity to health facilities, access to water sources, agricultural dependence and other community-level factors.

Preliminary findings suggested that regions with stronger infrastructure, better access to services, and higher levels of disaster preparedness tended to show faster rates of return and more stable population recovery. Additionally, socio-demographic characteristics and the presence of local social support networks appeared to play an important role in facilitating resilience. The group's work highlighted the potential of integrating digital and traditional data sources to uncover complex patterns of recovery and inform more effective post-disaster planning and resilience-building strategies.

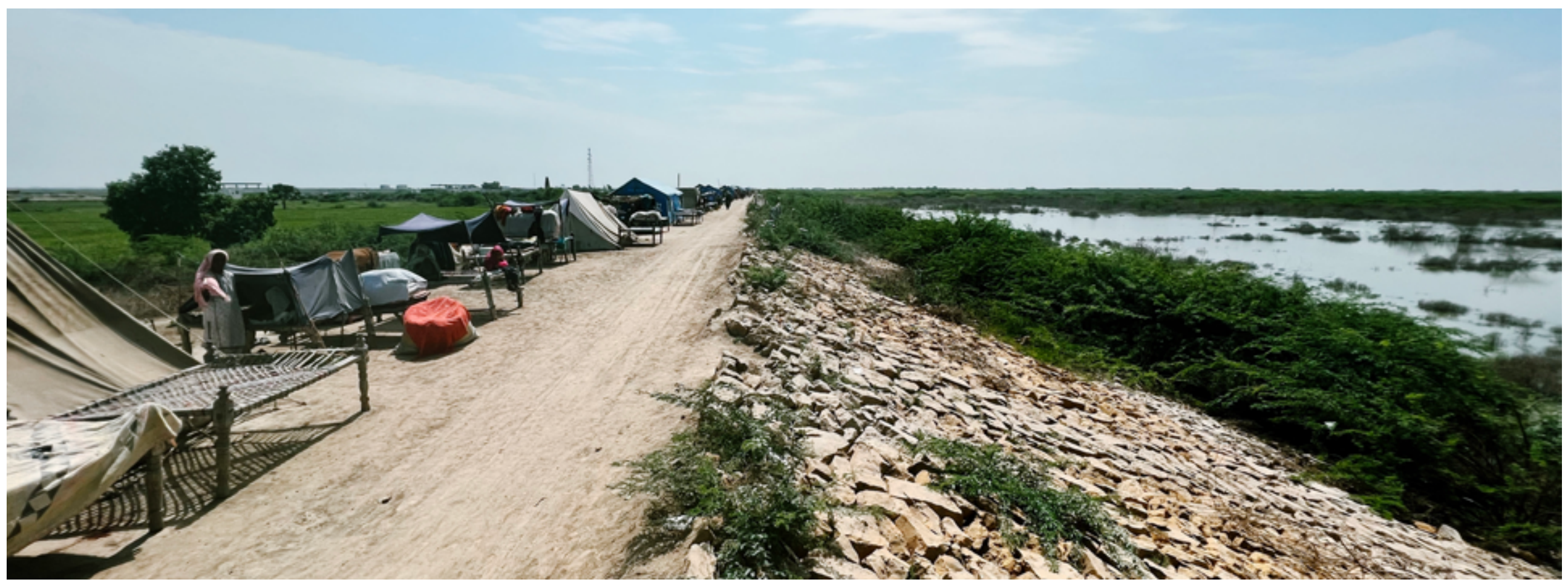

An IDP site stretching along elevated ground near the Thatta Sujawal road, south of Hyderabad, Sindh, Pakistan. © IOM 2022/Brian MCDONALD

**Group 4:** Carmen Cabrera, PhD (UoL), Matt Mason (UoL), Yaroslav Smirnov (IOM), Brian McDonald (IOM), Abdul Samad Omari (IOM), Dirk Jung (Snowflake), Alex McCarthy (Snowflake) and Rachel Cribbin (FCDO).

**Problem statement: Event-based or rapid response: Developing key indicators**

Group 4 addressed the challenge of "Event-based or rapid response: Developing key indicators," with the goal of creating timely, actionable insights to support immediate disaster response. A key focus was to display these indicators in an interactive, user-friendly dashboard suitable for operational use by humanitarian organizations.

The team began by developing a metric of population inflows using Meta's Facebook Movement During Crisis data at the 800 metres tile level. These data were aggregated to PCode geographies commonly used by humanitarian actors, ensuring compatibility with traditional datasets such as IOM's Displacement Tracking Matrix. This enabled the creation of an interactive map showing movement across Pakistan throughout the crisis period.

A central component of the analysis involved combining Facebook mobility data with high-resolution spatial data on deprivation. By linking inflow volumes with levels of deprivation, the team produced a bivariate map that highlighted areas experiencing both high population inflows and high deprivation – flagging them as likely hotspots of vulnerability where incoming and resident populations could be under strain.

To assess the representativeness of the digital nontraditional data, the team also calculated the proportion of Facebook users (as captured by the dataset) relative to the total population in each administrative unit. This helped highlight areas where findings from digital nontraditional data might be reliable. These estimates were mapped alongside the locations of Community Needs Initiative (CNI) data-collection sites, helping identify where digital trace data could complement or fill gaps in traditional field data.

Additionally, the team conducted a focused case study on Mirpur Khas in Sindh. Here, they used Facebook data to assess population changes during the floods and overlaid this with CNI data on community needs. Although time constraints limited further development, this case study served as a template for how digital and traditional data could be synthesized at a local level.

All outputs were compiled into an interactive dashboard, designed for accessibility by humanitarian responders. The dashboard can be accessed at: https://hackathon.humanitarian.im/#. This work illustrated the potential of integrating digital and traditional data to generate real-time, operational insights for disaster response.

**Figure 13. Screenshot of an interactive dashboard made by Group 4 during the hackathon**

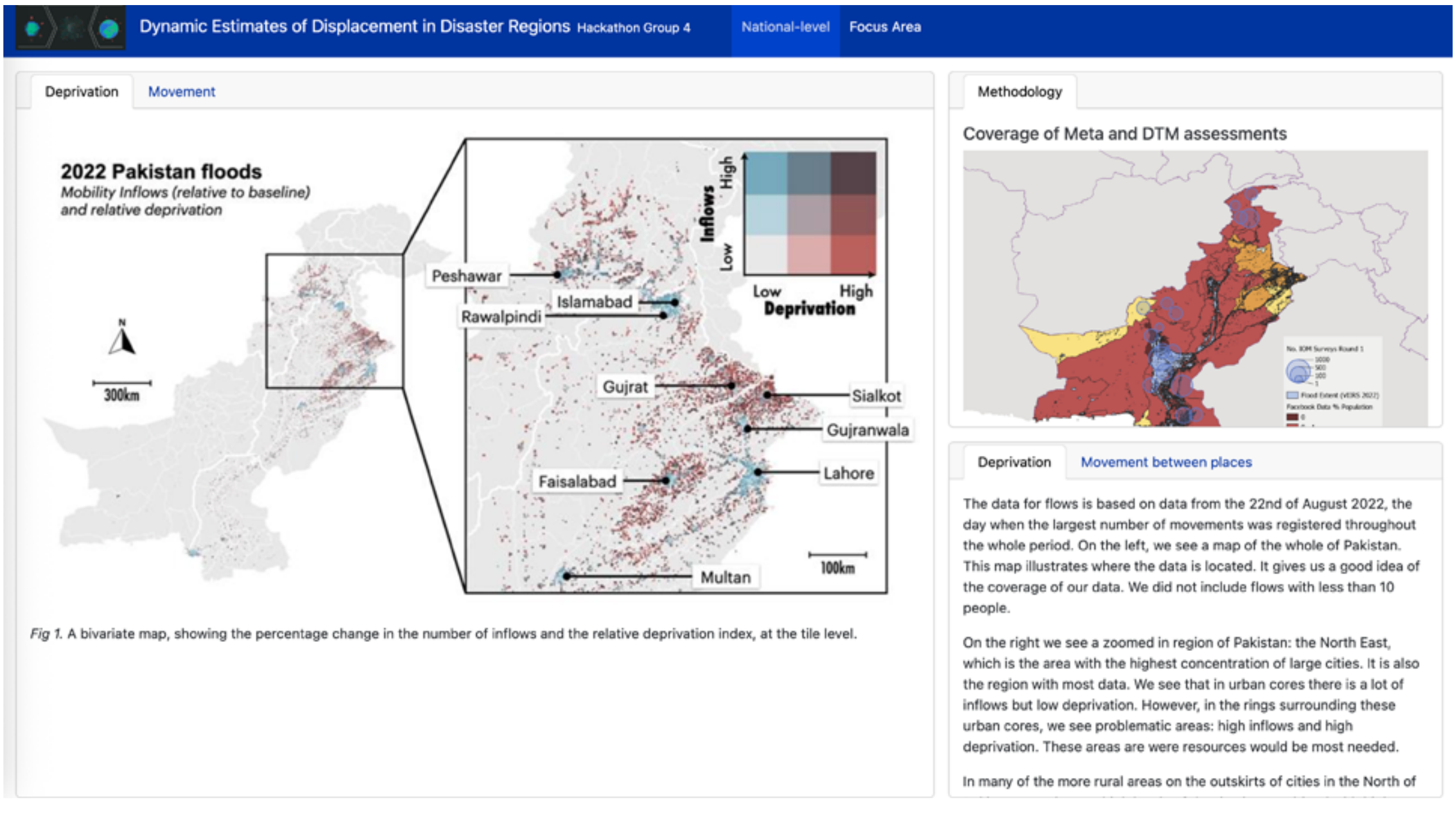

*Source*: Generated by Group 4 using Power BI during the Berlin hackathon and is not a final output but demonstrates work done during hackathon.
*Notes*: The left-hand map in the dashboard shows the flows from the Facebook Movement During Crisis, with a slider to change the time period the map displays. The right-hand map shows the areas of Pakistan flooded, the locations where the first round of the CNI collected data from, and the number of users captured by the Facebook data by administrative area as a percentage of the resident population. The maps are for illustration purposes only. The boundaries and names shown and the designations used on this map do not imply official endorsement or acceptance by the International Organization for Migration.

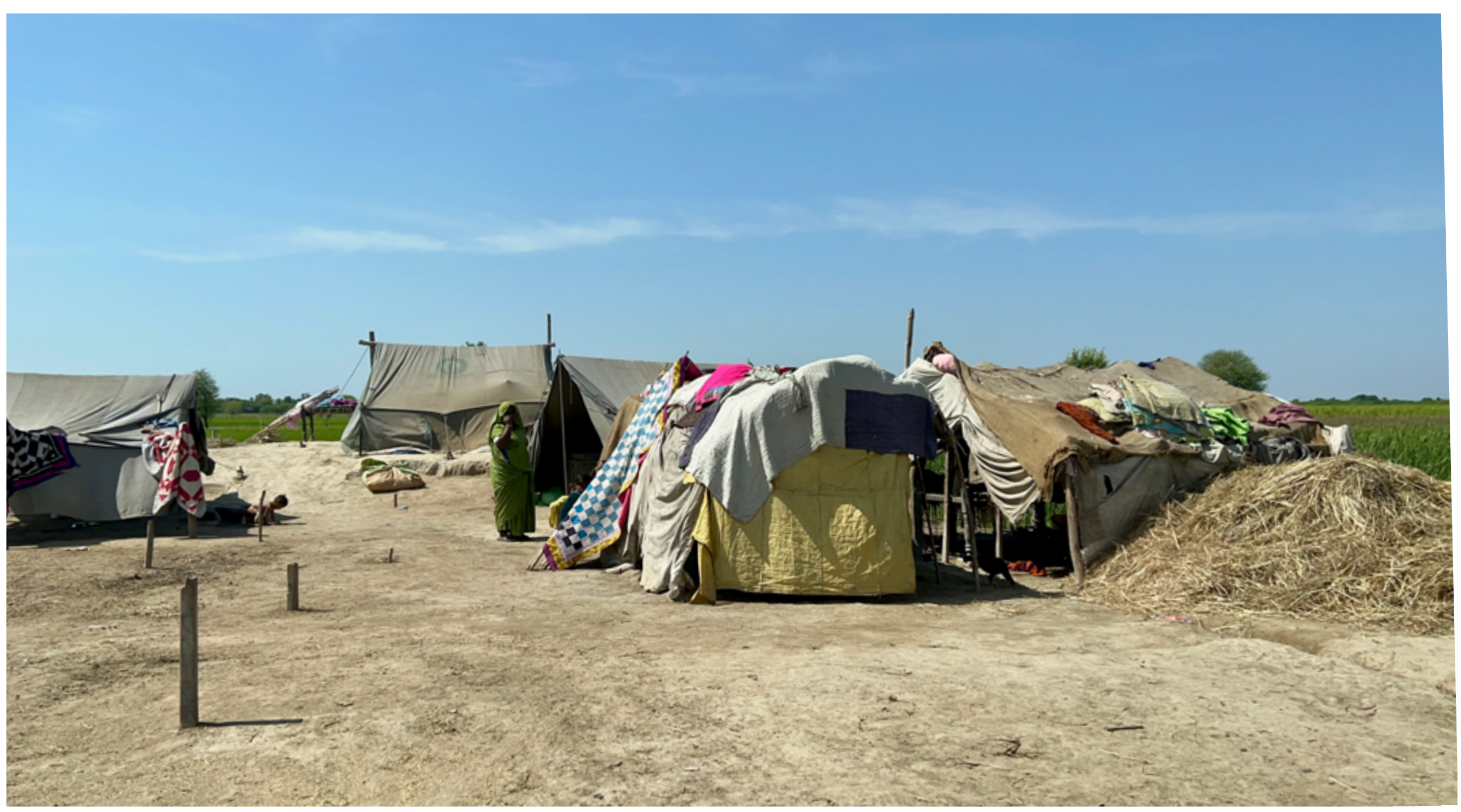

Flood affected community near Badin, Sindh, Pakistan. © IOM 2022/Brian MCDONALD

The hackathon's outputs demonstrate how digital trace data can add to the insights gained from traditional forms of data collection and how it can help policymakers and humanitarian response to disasters. Given the time constraints of the hackathon (the event lasted around a day and a half), each group's outputs were not fully developed. The progress that was made within a short period, however, shows the possibility for the successful integration of digital and traditional data going forward.

# CONCLUSION

Digital nontraditional data sources offer valuable alternatives in data-scarce environments, helping to fill critical information gaps during humanitarian crises. While digital trace data cannot replace the rich contextual insights provided by traditional sources – such as surveys, administrative records and field assessments – they serve as a powerful complement. This report has explored the potential of such approaches through two case studies: the escalation of the war in Ukraine (2022) and the Pakistan floods (2022), each illustrating their application in measuring displacement following major disasters. When used together, digital nontraditional and traditional data streams strengthen the overall resilience and reliability of the humanitarian data ecosystem.

This layered approach reduces reliance on any single data source, enabling more adaptive, comprehensive and timely insights to inform response efforts – particularly in contexts with constrained information access or limited resources. As the sustainability of many traditional data systems becomes increasingly uncertain, integrating digital sources offers a key opportunity to reinforce humanitarian data capacity and foster collaboration across sectors, including with data providers and researchers.

To maximize the potential of digital nontraditional data in humanitarian contexts, a coordinated approach is essential – one that prioritizes standardization, integration, timeliness, transparency, ethics and trust. Clear and consistent data definitions, alignment with existing humanitarian systems and scalable methodologies enhance usability and comparability across contexts. Integrating nontraditional digital sources into established data ecosystems can fill critical information gaps, especially in crisis onset or resource-constrained settings. Moreover, value lies not in raw data but in timely, actionable and responsibly processed outputs. This requires transparent, well-documented methodologies, ethical data engineering practices and clear communication of limitations. Building trust through robust governance frameworks, collaborative data-sharing protocols and trusted intermediaries is equally vital. Ultimately, by fostering an ecosystem of responsible innovation, digital nontraditional data can support faster, more informed and more equitable humanitarian responses.

The University of Liverpool and IOM DTM are committed to advancing this work, continuing to develop data triangulation approaches across diverse displacement contexts. With the frequency of both conflict-driven and climate-induced disasters on the rise, strengthening these methods is critical to better prepare for future crises. To embed and scale this project further, additional funding is sought to support the creation of an open-source package, designed to enhance usability, reproducibility and accessibility of these methodologies for practitioners and researchers alike.

* All hyperlinks were working at the time of writing this report.

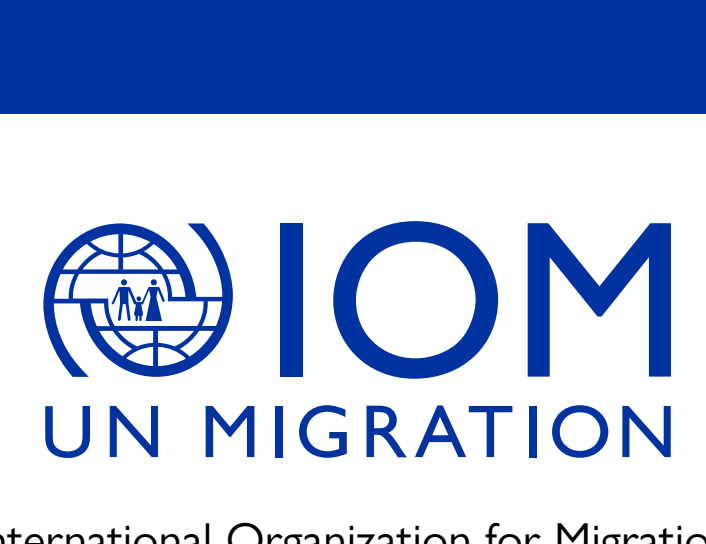

International Organization for Migration
17 route des Morillons, P.O. Box 17, 1211 Geneva 19, Switzerland
Tel.: +41 22 717 9111 • Fax: +41 22 798 6150 • Email: hq@iom.int • Website: www.iom.int